\documentclass[sn-mathphys-ay]{sn-jnl}

\usepackage{graphicx}%
\usepackage{multirow}%
\usepackage{amsmath,amssymb,amsfonts}%
\usepackage{amsthm}%
\usepackage{mathrsfs}%
\usepackage[title]{appendix}%
\usepackage{xcolor}%
\usepackage{textcomp}%
\usepackage{manyfoot}%
\usepackage{booktabs}%
\usepackage{algorithm}%
\usepackage{algorithmicx}%
\usepackage{algpseudocode}%
\usepackage{listings}%
\usepackage{tabularx} 

\theoremstyle{thmstyleone}%
\theoremstyle{thmstyletwo}%

\theoremstyle{thmstylethree}%

\begin{document}

\title[Article Title]{Deep Learning Model Inversion Attacks and Defenses: A Comprehensive Survey}


\author*[1]{\fnm{Wencheng} \sur{Yang}}\email{wencheng.yang@unisq.edu.au}
\author[2]{\fnm{Song} \sur{Wang}}\email{song.wang@latrobe.edu.au}
\author[1]{\fnm{Di} \sur{Wu}}\email{di.wu@unisq.edu.au}
\author[1]{\fnm{Taotao} \sur{Cai}}\email{taotao.cai@unisq.edu.au}
\author[3]{\fnm{Yanming} \sur{Zhu}}\email{yanming.zhu@griffith.edu.au}
\author[1]{\fnm{Shicheng} \sur{Wei}}\email{shicheng.wei@unisq.edu.au}
\author[4]{\fnm{Yiying} \sur{Zhang}}\email{yiyingzhang@tust.edu.cn}
\author[5]{\fnm{Xu} \sur{Yang}}\email{xu.yang@mju.edu.cn}
\author[1]{\fnm{Zhaohui} \sur{Tang}}\email{zhaohui.tang@unisq.edu.au}
\author[1]{\fnm{Yan} \sur{Li}}\email{yan.li@unisq.edu.au}



\affil[1]{\orgname{University of Southern Queensland}, \orgaddress{\city{Toowoomba}, \postcode{4350}, \state{QLD}, \country{Australia}}}

\affil[2]{\orgname{La Trobe University}, \orgaddress{\city{Melbourne}, \postcode{3083}, \state{VIC}, \country{Australia}}}

\affil[3]{\orgname{Griffith University}, \orgaddress{\city{Gold Coast}, \postcode{4222}, \state{QLD}, \country{Australia}}}

\affil[4]{\orgname{Tianjin University of Science and Technology}, \orgaddress{\city{Tianjin}, \postcode{300222}, \country{China}}}

\affil[5]{\orgname{Minjiang University}, \orgaddress{\city{Fuzhou}, \postcode{350108}, \state{Fujian}, \country{China}}}


\abstract{The rapid adoption of deep learning in sensitive domains has brought tremendous benefits. However, this widespread adoption has also given rise to serious vulnerabilities, particularly model inversion~(MI) attacks, posing a significant threat to the privacy and integrity of personal data. The increasing prevalence of these attacks in applications such as biometrics, healthcare, and finance has created an urgent need to understand their mechanisms, impacts, and defense methods. This survey aims to fill the gap in the literature by providing a structured and in-depth review of MI attacks and defense strategies. Our contributions include a systematic taxonomy of MI attacks, extensive research on attack techniques and defense mechanisms, and a discussion about the challenges and future research directions in this evolving field. By exploring the technical and ethical implications of MI attacks, this survey aims to offer insights into the impact of AI-powered systems on privacy, security, and trust. 
In conjunction with this survey, we have developed a comprehensive repository to support research on MI attacks and defenses. The repository includes state-of-the-art research papers, datasets, evaluation metrics, and other resources to meet the needs of both novice and experienced researchers interested in MI attacks and defenses, as well as the broader field of AI security and privacy. The repository will be continuously maintained to ensure its relevance and utility. It is accessible at https://github.com/overgter/Deep-Learning-Model-Inversion-Attacks-and-Defenses.}




\maketitle

\section{Introduction}

As deep learning models are increasingly integrated into sensitive applications, robust privacy and security measures are critical. While the deep learning mechanism provides unprecedented predictive accuracy, it also introduces vulnerabilities related to how models encode and retain information about training data ~\citep{Rigaki_Garcia_2023, Sen_Waghela_Rakshit_2024}. Complex architectures, such as convolutional neural networks (CNNs) and transformers, are often effective at generalization, but may inadvertently remember details of training data. In privacy-sensitive applications, such memorization poses a significant risk, as indirect inferences about training data can lead to serious consequences.

\begin{figure}[h!]
  \centering
  \includegraphics[width=\linewidth]{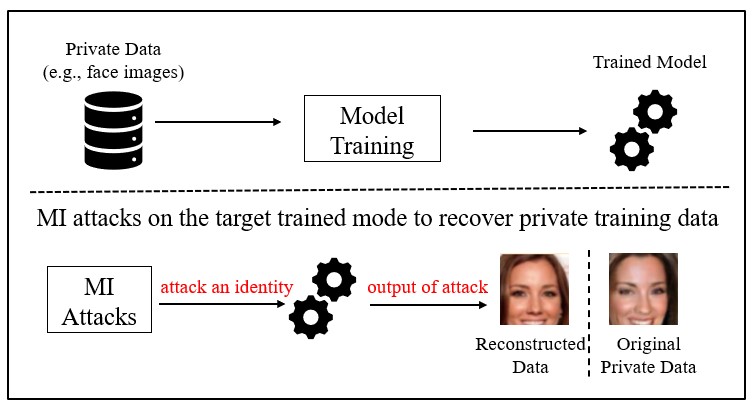}
  \caption{A visualization example of the model inversion (MI) attack.}
  \label{fig_MIA_example}
\end{figure}

Among major attacks that threaten deep learning systems (e.g., model extraction, model inversion (MI), poisoning, and adversarial attacks)~\citep{He_Meng_Chen_Hu_He_2020}, the MI attack stands out due to its ability to extract sensitive information from the training dataset and compromise user privacy~\citep{Gong_Jiang_Liu_Wang_Gastro_Wang_Zhang_Guo_2023}. MI attacks were proposed by Fredrikson et al.~\citep{Fredrikson_Lantz_Jha_Lin_Page_Ristenpart_2014}, which exploits the relationship between input data and model learning parameters to recover sensitive information. The impact of MI attacks goes beyond violations of personal privacy, undermining trust in machine/deep learning systems in critical applications such as biometrics, healthcare analytics, and financial modeling. For example, using MI attacks on face recognition systems, an adversary can approximate face features in the training dataset and reconstruct the face image, thereby undermining the security of biometric systems~\citep{Yang_Wang_Cui_Tang_Li_2023, Tran_Nguyen_Mai_Vandierendonck_Cheung_2024, Qiu_Zhang_Ji_Fu_Yang_Wang_2024}. In healthcare systems, leaked sensitive patient data may violate ethical standards and lead to legal consequences~\citep{Dao_Nguyen_2024, Tang_Van_Nguyen_Yang_Xia_Chen_Mullens_Dean_Osborne_Li_2024, Nguyen_Yang_Tang_Xia_Mullens_Dean_Li_2024}. Similarly, in financial systems, private transaction data or credit scores could be compromised by MI attacks~\citep{Milner_2024}. In addition to privacy concerns, these attacks can erode users' trust in machine/deep learning systems, as users may lose confidence in the security of their personal or sensitive information. A visualization example of the MI attack is shown in Figure~\ref{fig_MIA_example}
(adapted from~\citep{Nguyen_Chandrasegaran_Abdollahzadeh_Cheung_2023}).

From a technical perspective, MI attacks exploit the inherent vulnerability of deep learning models to overfit or memorize patterns in training data ~\citep{Titcombe_Hall_Papadopoulos_Romanini_2021}. Researchers have focused on understanding the factors that exacerbate this vulnerability and developing robust defenses. There are several factors that affect the success of MI attacks~\citep{Fredrikson_Jha_Ristenpart_2015, Shokri_Stronati_Song_Shmatikov_2017}. First, access to the model: the attacker needs to access the model's outputs, architecture, or parameters. In white-box scenarios, the attacker has full knowledge of the model, including its architecture and parameters, while in black-box scenarios, only the output can be accessed. Second, model architecture: different models retain different levels of information about training data. Complex models, such as deep neural networks~(DNNs), are particularly vulnerable to MI attacks because they are more likely to memorize and store detailed information about their inputs. Third, data correlation: MI attacks typically exploit the correlation between input data and model output. By understanding how small changes in inputs affect model outputs, an adversary can fine-tune the reconstruction of original data.

Defending against MI attacks presents a unique challenge. First, designing a defense strategy requires balancing the utility and predictive power of the model with risk-reducing privacy protection mechanisms. Second, MI attacks usually require only limited information, such as access to model outputs or intermediate features, rather than full access to model parameters or training data~\citep{Han_Choi_Lee_Kim_2023, Fang_Chen_Wang_Wang_Xia_2023}. Therefore, preventing them is non-trivial. In addition, privacy-preserving techniques tend to degrade model performance, thus requiring a careful trade-off between accuracy and privacy.

MI attacks highlight a critical vulnerability in deep learning models, emphasizing the need for strong safeguards to protect sensitive information. This survey explores the taxonomy of MI attacks and examines the defense strategies proposed to mitigate these threats.

\subsection{Existing Surveys on MI Attacks and Defenses}

A number of existing surveys on MI attacks and defenses provide valuable insights into the mechanisms, challenges, and countermeasures associated with this privacy threat. These surveys are summarized below.

Zhang et al.~\citep{Zhang_Guo_Wang_Xie_Tao_2022} found that training samples can be reconstructed from gradients by techniques known as gradient inversion (GradInv) attacks. These attacks show how attackers can exploit gradient data to endanger data privacy. The authors categorized GradInv attacks into two main modes (i.e., paradigm-iteration-based and recursion-based approaches), highlighting key techniques for gradient matching and data initialization to optimise recovery. Dibbo~\citep{Dibbo_2023} presented a systematic review of MI attacks, providing a taxonomy to classify MI attacks based on a variety of methods and features. The taxonomy highlights the unique nature of MI attacks compared to other privacy threats (e.g., model extraction and membership inference), positioning MI attacks as attacks with unique complexity and impact. The author also outlined key defense strategies and identifies a number of open questions. Yang et al.~\citep{Yang_Ge_Xue_Xiang_Li_Lu_2023} provided an in-depth study of gradient leakage attacks in federated learning~(FL) and classified these attacks into optimisation-based and analysis-based attacks. The former considers data reconstruction as an optimisation problem, while the latter solves the problem through linear equation analysis. To address the limitations of these traditional approaches, the authors proposed a novel generation-based paradigm that greatly improves the accuracy and efficiency of reconstruction.

Fang et al.~\citep{Fang_Qiu_Yu_Yu_Kong_Chong_Chen_Wang_Xia_Xu_2024} provided a comprehensive survey of MI attacks and defenses, shedding light on different approaches and features used in DNN applications. Their study systematically categorizes MI attacks based on data types and target tasks, outlines early MI techniques in traditional machine/deep learning, and then shifts the focus to sophisticated modern approaches in DNNs. The authors also provided a defense taxonomy that explores current efforts to reduce MI risks. Liu et al.~\citep{Liu_Wang_Lei_2024} presented a systematic evaluation of data reconstruction attacks, criticizing previous studies for relying on empirical observations without sufficient theoretical foundations. The authors introduced an approach that enables a theoretical assessment of data leakage and establishes bounds on reconstruction error, especially for two-layer neural networks. This study identifies upper bounds on the reconstruction error and information theoretic lower bounds, advancing the understanding of attack and defense dynamics. 

Shi et al.~\citep{Shi_Kotevska_Reshniak_Singh_Raskar_2024} conducted a survey on gradient inversion attacks (GIAs) in FL that addresses the critical need to expand from the traditional “honest but curious” server model to a threat model involving malicious servers and clients. The authors classified GIAs according to the roles of these adversaries, demonstrating that traditional defenses are often ineffective against more aggressive malicious actors. The authors also  detailed various attack strategies in their classification, specifically how malicious servers and clients can bypass existing defenses, and highlighted gaps in FL's ability to deal with such complex threats. The work stresses the role of reconstruction methods, model architectures, and evaluation metrics in shaping the effectiveness of defenses. This research underscores the importance of developing stronger defenses against GIAs.



\textbf{Differences to Existing Surveys.} The main differences between our work and existing surveys can be summarized as follows. First, although the existing surveys have advanced the categorization of MI attacks and their impact, these reviews lack a unifying framework that integrates the latest advances across various paradigms and applications. Second, they fail to provide a comprehensive summary of emerging challenges and corresponding future research directions. Third, many of these studies narrowly focus on specific techniques (e.g., only about gradient inversion attacks and defenses), leaving critical gaps in understanding broader MI attacks and defenses.

Our survey aims to address aforementioned limitations through a comprehensive and integrated analysis of MI attacks, defenses and other aspects (e.g., applications, data types, and evaluation metrics). By combining theoretical insights with practical considerations in different applications, our work seeks to advance research on MI defense and guide the development of robust privacy protection strategies. While almost all MI-related surveys cover some common topics such as attacks, defenses, and future research, our work provides a more in-depth review as it explores additional aspects that are not covered by existing surveys, such as different recommendations regarding future research directions. Moreover, we create a unified resource repository for studying MI attacks and defenses. This will benefit both novice and experienced researchers in the field of deep learning-related security and privacy. A comparison between existing surveys and our work is presented in Table~\ref{table_survey_comparison}.

\begin{table}[h!]
\centering
\renewcommand{\arraystretch}{1.5} 
\setlength{\tabcolsep}{6pt} 
\resizebox{\textwidth}{!}{
\begin{tabular}{|m{1.5cm}<{\centering}|m{1.5cm}<{\centering}|m{1.5cm}<{\centering}|m{1.5cm}<{\centering}|m{1.5cm}<{\centering}|m{1.5cm}<{\centering}|m{1.5cm}<{\centering}|m{1.5cm}<{\centering}|m{1.5cm}<{\centering}|m{1.5cm}<{\centering}|}
\hline
\textbf{Surveys} &\textbf{Gradient Inversion Attacks} &\textbf{Generative Model-based Attacks} &\textbf{Different Applications} &\textbf{Different Data Types} &\textbf{Defenses} &\textbf{Evaluation Metrics} &\textbf{Datasets} &\textbf{Challenges and Future Research} &\textbf{Resource Repository}  \\ \hline
\citep{Zhang_Guo_Wang_Xie_Tao_2022}& Yes & No & No & No & Yes & No & No & Yes & No          \\ \hline
\citep{Dibbo_2023} & Yes & Yes & Yes & No & Yes & No & No & Yes & No\\ \hline
\citep{Yang_Ge_Xue_Xiang_Li_Lu_2023} & Yes & No & No & No & No & No & No  & Yes & No\\ \hline
\citep{Fang_Qiu_Yu_Yu_Kong_Chong_Chen_Wang_Xia_Xu_2024}  & Yes & Yes & No & Yes & Yes & No & No & Yes & Yes \\ \hline
\citep{Liu_Wang_Lei_2024} & Yes & Yes & No & No & Yes & No & No & No & No \\ \hline
\citep{Shi_Kotevska_Reshniak_Singh_Raskar_2024} & Yes & No & No & No & Yes & No & No & Yes & No \\ \hline
\textbf{This Survey} & Yes & Yes& Yes & Yes& Yes & Yes& Yes & Yes& Yes\\ \hline
\end{tabular}
}
\caption{A comparison of existing surveys and this work. In this table, ``Yes" indicates that the topic is covered, while ``No" means that it is not covered.}
\label{table_survey_comparison}
\end{table}

\subsection{Contributions of This Work}

The main contributions of this research are summarized below.

\textbf{The first major contribution of this work is the development of a new, structured taxonomy stemming from diverse techniques proposed in recent years.} The new taxonomy provides a clear, unified framework for understanding and categorizing MI attacks based on key factors, such as methodology (e.g., gradient-based, generative model-based, and optimisation-based), data type (e.g., image, audio, text/tabular data), and application domain (e.g., biometric recognition, healthcare, finance). This structured organization not only enhances the understanding of the differences and vulnerabilities of MI attacks in various scenarios, but also highlights research gaps and lays the groundwork for exploring defense strategies. Complementing the taxonomy, the comprehensive review systematically analyzes the state of the art, including gradient inversion, generative model-based, and optimisation-based attacks. Detailed case studies and comparisons of different strategies provide insight into the impact, limitations, and practical applications of these strategies. The taxonomy and review can serve as a valuable resource for researchers and practitioners to guide future research and defense against MI attacks.

\textbf{The second contribution is a more in-depth and extensive review of defense mechanisms designed to mitigate MI attacks than existing surveys.} While this survey investigates and analyzes current development and existing approaches, specific ideas are also studied. The main defense mechanisms include feature/gradient perturbation, i.e., concealing or obscuring sensitive information (e.g., feature representations and gradients) to distort the data and thwart data reconstruction. Differential privacy, where noise is injected into the training process to protect sensitive data; and cryptographic encryption, which allows training the model using encrypted data without exposing the original input. By analyzing these mechanisms, this survey provides a roadmap for strengthening models against MI attacks while maintaining effectiveness in real-world applications.

\textbf{The third contribution is to identify open questions and suggest future research directions}. Despite the progress in understanding and mitigating MI attacks, critical challenges remain. These challenges include developing defenses that can balance privacy protection with model utility, improving the robustness of deep learning models against increasingly sophisticated MI attacks, and addressing vulnerabilities in emerging paradigms such as FL and edge AI systems. This survey highlights these challenges and outlines viable research directions to stimulate innovative research in defending MI attacks.

\textbf{The final contribution is the creation of a unified resource repository for studying MI attacks and defenses.} We compile and standardize the name abbreviations of existing MI attack and defense methods. In cases where the authors do not provide abbreviations, we generate reasonable abbreviations based on our understanding to ensure clarity and usability. These standardized abbreviations help researchers utilize these works in their own research. In addition, we compile a list of datasets and evaluation metrics commonly used in MI attack and defense research. To consolidate these resources, we develop a comprehensive repository of state-of-the-art research articles, datasets, evaluation metrics, and other important resources. This repository is intended to support both novice and experienced researchers in the study of MI attacks, defenses, and the broader field of AI security and privacy. It will be continually maintained to ensure its relevance and accessibility. The repository is available at https://github.com/overgter/Deep-Learning-Model-Inversion-Attacks-and-Defenses.

\subsection{Literature Selection Methodology}
This study employs a systematic method to identify, analyze, and synthesize relevant research articles on deep learning model inversion attacks and defenses.
A thorough search is conducted across multiple reputable digital libraries, including IEEE Xplore, ACM Digital Library, SpringerLink, ScienceDirect, arXiv, and top-tier conferences such as CVPR, NeurIPS, ICML, ECCV, USENIX, and NDSS. The literature search covers foundational studies as well as latest advancements in the areas of interest. Both journal and conference papers, as well as preprints and technical reports, are included so as to achieve a complete coverage and a comprehensive overview.

To retrieve relevant studies, Boolean search queries are formulated with primary keywords, including ``Model Inversion Attack", ``Gradient Inversion Attack", ``Privacy Breach in Deep Learning", ``Federated Learning Security", ``Privacy-Preserving AI", ``Generative Model Inversion", ``GAN-based Model Inversion", and ``Data Reconstruction Attack". The Population, Concept, and Context framework~\citep{Ullah_Manickam_Obaidat_Laghari_Uddin_2023} is applied to refine the search scope. The population consists of deep learning architectures, including CNNs, RNNs, transformers, and FL models. The concept focuses on techniques and defenses related to MI attacks, while the context covers AI security, biometric recognition, healthcare privacy, and cloud-based AI services. 

A structured filtering process is implemented using Zotero 7.0 for reference management and duplicate removal. The inclusion criteria ensure the selection of peer-reviewed journal and conference papers on MI attacks and defenses, studies introducing novel attack techniques such as GAN-based, black-box, and gradient inversion attacks, experimental research evaluating attack effectiveness and defense robustness, and review papers summarizing up-to-date developments on topics pertaining to deep learning MI attacks and defenses. The exclusion criteria eliminate studies lacking empirical validation or theoretical discussions without experimental results, redundant works, and short papers, editorials, and opinion articles without technical depth. As a result, about 500 research articles are retrieved from various sources. Following a careful assessment to titles, abstracts, and full-text review, it results in a final selection of approximately 180 high-quality papers for this survey. The selected articles are collated into four themes. The first theme is about MI attack methodologies. The second theme explores defensive countermeasures and examination methods. The third theme focuses on evaluation metrics and datasets for MI attack research. The fourth theme addresses emerging challenges and future research directions.


\section{Fundamentals of MI Attacks}
\label{sec:Fundamentals}
MI attacks pose a significant threat to the privacy and security of deep learning models by attempting to reconstruct sensitive input data from model parameters, outputs, or intermediate representations. Understanding the underlying mechanisms of MI attacks is crucial for designing robust defense strategies. This section provides a comprehensive overview of MI attacks, beginning with their definition and characteristics, followed by an analysis of how these attacks exploit different components of deep learning models under various adversarial knowledge scenarios.
\subsection{Definition and Characteristics}
This subsection covers the mathematical formulation of deep learning models, their optimization process, and MI attacks that extract sensitive data from model parameters, outputs, or representations.
\subsubsection{Deep Learning Models}
As a subset of machine learning, deep learning employs artificial neural networks with multiple layers to model complex patterns in data. A deep learning model is composed of interconnected layers of neurons, where each layer applies mathematical transformations to input data to learn representations that are progressively more abstract~\citep{Goodfellow_2016}. The operation of a single layer can be mathematically described as

\begin{equation}
h^{(l)} = f\left(W^{(l)} \cdot h^{(l-1)} + b^{(l)}\right)
\end{equation}
where \( h^{(l-1)} \) is the input from layer~$l-1$ (i.e., previous layer);  \( W^{(l)} \) is the weight matrix of the \( l \)th layer; \( b^{(l)} \) is the bias vector of the \( l \)th layer; \( f(\cdot) \) is the activation function (e.g., ReLU, sigmoid); and \( h^{(l)} \) is the output of layer~$l$ (i.e., current layer). For a network with multiple layers (e.g., \(n\) layers), the final prediction is

\begin{equation}
\hat{y} = \sigma\left(f_n \circ f_{n-1} \circ \ldots \circ f_1(\mathbf{x})\right)
\end{equation}
where symbol~\(\circ\) denotes function composition, which means that the output of one function becomes the input to the next function; and  \( \sigma(\cdot) \) is a decision function (e.g., softmax) for classification.

\textbf{Model Parameters.} Given a deep learning model, weight~(\(W\)) represents the connection strength between neurons in adjacent layers and is updated iteratively to improve the model's predictions. The loss function \( \ell \) quantifies the discrepancy between the model’s predictions and the actual target values, guiding the optimization process. Gradient \( \nabla W \) can be expressed by
\begin{equation} \label{gradient}
\nabla W = \frac{\partial \ell}{\partial W}
\end{equation}
where \( \nabla W \) is the gradient of the weight matrix~\( W \), indicating the direction and magnitude of change needed to minimize the loss function \( \ell \)~\citep{Zhu_Liu_Han_2019}. The term~\( \frac{\partial \ell}{\partial W} \) in Equation~(\ref{gradient}) denotes the partial derivative of the loss function \( \ell \) with respect to the weight matrix~\( W \), capturing how small changes in \( W \) impact the loss.

\subsubsection{MI Attacks on Deep Learning Models}

MI attacks exploit a deep learning model’s parameters (e.g., gradients~\citep{Zhu_Liu_Han_2019, Zhao_Mopuri_Bilen_2020}), output (e.g., confidence scores~\citep{Han_Choi_Lee_Kim_2023}) or intermediate representations~\citep{Fang_Chen_Wang_Wang_Xia_2023} to reconstruct sensitive information about its inputs (e.g., training data). Depending on the specific component being targeted, the optimisation process for MI attacks varies:

\textbf{MI Attacks Using Parameters (e.g., gradients).} The process involves solving an optimisation problem, typically as follows (quoted from \citep{Zhu_Liu_Han_2019}):
\begin{equation}
(\mathbf{x}'^*, \mathbf{y}'^*) = \arg \min_{\mathbf{x}', \mathbf{y}'} \|\nabla W' - \nabla W\|^2 = \arg \min_{\mathbf{x}', \mathbf{y}'} \left\| \frac{\partial \ell(F(\mathbf{x}', W), \mathbf{y}')}{\partial W} - \nabla W \right\|^2
\end{equation}
where \(\mathbf{x}'^*\) and \(\mathbf{y}'^*\) are the recovered (dummy) input and label, which are iteratively optimised to minimize the difference between the dummy gradients and the real gradients; \(\nabla W\) represents the real gradients of the loss function with respect to the model weights \(W\), computed using the true training data \((\mathbf{x}, \mathbf{y})\); \(\nabla W'\) represents the dummy gradients, computed using the dummy data \((\mathbf{x}', \mathbf{y}')\); and \(\|\cdot\|^2\) is the objective function that aims to minimize the squared difference (L2 norm) between the dummy gradients \(\nabla W'\) and the real gradients \(\nabla W\). By minimizing this difference, the dummy data becomes increasingly similar to the true training data.

\textbf{MI Attacks Using Outputs or Intermediate Representations.} These attacks rely on extracting meaningful information directly from confidence scores or representations of intermediate layers, often without requiring explicit optimisation~\citep{Han_Choi_Lee_Kim_2023, Fang_Chen_Wang_Wang_Xia_2023}. For instance, intermediate representations may be analyzed to reconstruct input features without the iterative process used for the MI attacks using parameters.

\subsection{MI Attacks under Different Model Knowledge Scenarios}
Under different model knowledge scenarios, MI attacks may be carried out in different ways; that is, access to the target model may be white-box, black-box or gray-box for adversaries~\citep{Pengcheng_Yi_Zhang_2018, Dibbo_2023}.

\textbf{White-box Access.} The adversary has full access to the model's architecture and training algorithm. The adversary also has access to parameters (e.g., weights) or middle layer activations.

\textbf{Black-box Access.} The adversary interacts with the model only through the application programming interface~(API), providing input queries and observing corresponding outputs, such as prediction labels, class probabilities (softmax outputs), and logits (unnormalized predictions). The adversary does not have direct access to the model's parameters.

\textbf{Gray-box Access.} It is also considered as a semi-white scenario~\citep{Khosravy_Nakamura_Hirose_Nitta_Babaguchi_2021}, in which the adversary has partial access, such as access to the architecture (e.g., model structure), but not access to the parameters or training details. In some cases, the adversary may know specific layers or output formats.

\section{Taxonomy of MI Attacks}
\label{sec:Taxonomy MIA}

\begin{figure}[h!]
  \centering
  \includegraphics[width=\linewidth]{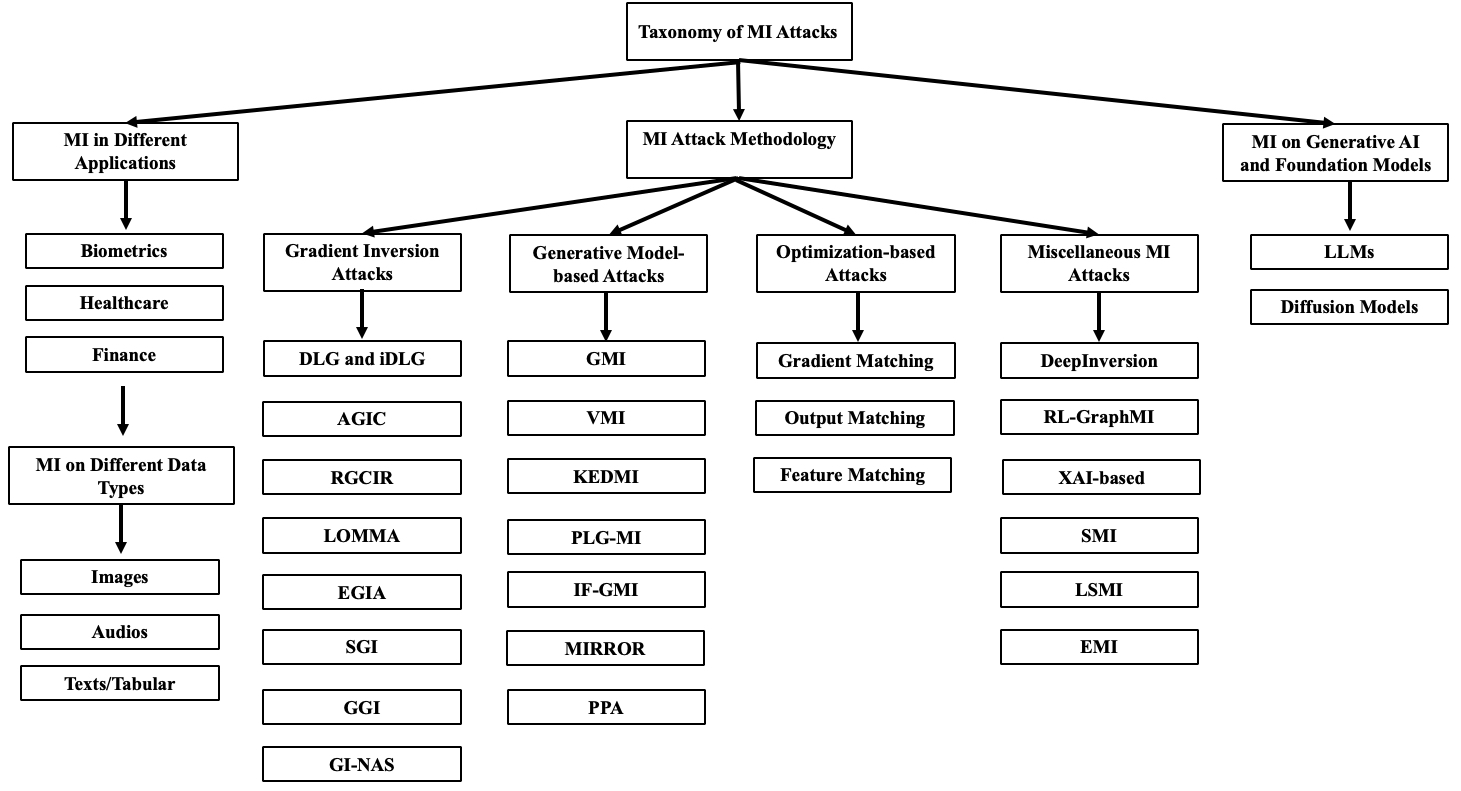}
  \caption{The structured taxonomy of the MI attacks in this survey.}
  \label{fig_MIA_taxonomy}
\end{figure}

MI attacks have become a major privacy threat that utilizes various methods to reconstruct sensitive data from machine/deep learning models~\citep{Pang_Chen_Deng_Wu_Bai_Xu_2024}. In this section, MI attacks are classified into three strategies: gradient inversion attacks, generative model-based attacks, and optimisation-based attacks, each of which has its own operational mechanisms and applications. The classification of MI attacks is illustrated in Figure~\ref{fig_MIA_taxonomy}. By exploring the unique features, strengths, and limitations of these strategies, we provide this structured taxonomy to facilitate a comprehensive understanding of the evolving MI attacks.

\subsection{Gradient Inversion Attacks} \label{gradient_inv}

\begin{figure}[h!]
  \centering
  \includegraphics[width=\linewidth]{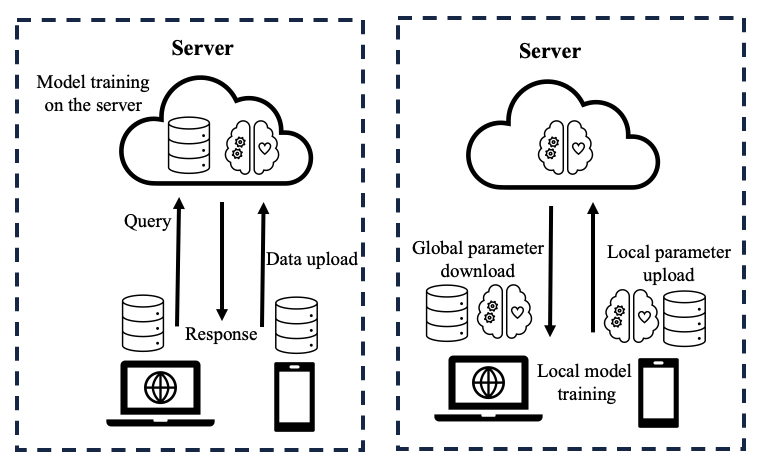}
  \caption{Illustration of the differences between centralized learning (left) and federated learning (right).}
  \label{Fig_CL_VS_FL}
\end{figure}

FL (federated learning) is a decentralized approach where model training occurs across multiple devices, exchanging only model parameters instead of raw data to enhance privacy. In contrast, centralized learning~(CL) collects all data on a central server for training purposes, offering efficiency but increasing privacy and security risks. The differences between FL and CL are demonstrated in Figure~\ref{Fig_CL_VS_FL} (adapted from~\citep{Dibbo_2023}).
FL is still vulnerable to gradient inversion attacks, which aim to reconstruct private data from shared gradients during training. In FL environments, attackers can exploit the gradients exchanged between clients and the central server to recover sensitive input data, undermining the privacy benefits of decentralization.

Geiping et al.~\citep{Geiping_Bauermeister_Dröge_Moeller_2020} conducted an in-depth investigation into the privacy risks in FL by analyzing the security of sharing parameter gradients. Contrary to the belief that gradient sharing is a privacy-preserving measure, the authors demonstrated that high-resolution images could be reconstructed from parameter gradients, endangering privacy. The authors gave a detailed analysis of how architectural choices and parameters could influence the reconstruction of input images. Although differential privacy and gradient compression were previously considered strong defense against gradient inversion attacks, Wu et al.~\citep{Wu_Chen_Guo_Weinberger_2023} demonstrated that a simple adaptive method, leveraging a model trained on auxiliary data, could successfully invert gradients and compromise the privacy of both vision and language tasks, even in the presence of these defenses, showing that existing countermeasures may have underestimated the privacy risks in FL.

\textbf{DLG and iDLG.} There are two basic gradient inversion attacks, called DLG~\citep{Zhu_Liu_Han_2019} and iDLG~\citep{Zhao_Mopuri_Bilen_2020}, which aim to reconstruct (steal) an FL client’s local data instances using the communicated $\nabla W$ gradients. The attacker generates a pair of dummy data \(\mathbf{x}'\) and dummy labels \(\mathbf{y}'\), which are used to generate dummy gradients (or weight update) $\nabla W^{\prime}$. Then by optimising the dummy gradients $\nabla W^{\prime}$ to be close to the real gradients $\nabla W$, the dummy data \(\mathbf{x}'\) will get close to the real data \(\mathbf{x}\). A key difference between DLG and its improved version iDLG lies in the way they extract the ground truth labels~\citep{Palihawadana_Wiratunga_Kalutarage_Wijekoon_2023}.

\textbf{The Approximate Gradient Inversion Attack (AGIC).} Xu et al.~\citep{Xu_Hong_Huang_Chen_Decouchant_2022} proposed AGIC, targeting FL models to reconstruct sensitive data through gradient or model updates. Unlike traditional attacks that focus solely on single mini-batch updates, AGIC introduces a more realistic scenario by considering multiple updates across several epochs. This approach estimates gradient updates to avoid expensive simulations and enhances reconstruction quality by assigning increasing weights to layers in the neural network structure. The authors demonstrated AGIC’s effectiveness through evaluations on CIFAR-10, CIFAR-100, and ImageNet datasets. The method achieved as much as a 50\% improvement in peak signal-to-noise ratio~(PSNR), while operating five times faster than earlier simulation-based attacks. 

\textbf{Refine Gradient Comparison and Input Regularization (RGCIR).} Developed by Luo et al.~\citep{Luo_Zhu_Fang_Kou_Hou_Wang_2022}, RGCIR is a more practical and effective method that addresses the limitations of existing gradient inversion techniques by overcoming prior constraints, such as dependency on batch normalization or limited applicability to small datasets. The authors utilized cosine similarity to refine gradient comparison and input regularization to improve image reconstruction fidelity. Additionally, the authors employed total variation denoising to enhance the smoothness of reconstructed images. Experimental evaluations demonstrate that RGCIR successfully recovers high-fidelity training data, even from large datasets like ImageNet, marking a significant advancement in the field. 

\textbf{Logit Maximization + Model Augmentation (LOMMA).} Nguyen et al.~\citep{Nguyen_Chandrasegaran_Abdollahzadeh_Cheung_2023} re-examined current MI techniques, uncovering two main issues that limit their performance: sub-optimal optimisation objectives and the problem of MI overfitting. The authors proposed LOMMA with a revised optimisation objective that significantly improves the performance of MI attacks, delivering notable enhancements across   state-of-the-art MI algorithms. Additionally, they introduced a segmentation-adaptive model augmentation strategy to address MI overfitting, a common issue where reconstructed images fail to accurately capture the semantics of the training data. 

\textbf{An External Gradient Inversion Attack (EGIA).} Liang et al.~\citep{Liang_Li_Zhang_Liu_Zhu_2023} introduced EGIA that operate in a gray-box setting and target public shared gradients transmitted through intermediate nodes in FL systems. Their work shows that even if both the client and the server are honest and fully trustworthy, an external adversary can still reconstruct private inputs by intercepting the transmitted gradients. The authors conducted extensive experiments on multiple real-world datasets and verify the effectiveness of EGIA. This work highlights the evolving threats in FL, where privacy vulnerabilities persist even in seemingly secure settings. 

\textbf{Stepwise Gradient Inversion (SGI). } Ye et al.~\citep{Ye_Luo_Zhou_Tang_2024} proposed SGI, a two-step approach that significantly improves the quality of image reconstruction. The authors first modeled the coefficient of variation (CV) of features, and then applied an evolutionary algorithm to accurately recover labels, followed by a stepwise gradient inversion attack to optimise the convergence of attack results. This method successfully recovers high-resolution images from complex models and large batch sizes, revealing the inherent vulnerabilities of distributed learning. 

\textbf{Generative Gradient Inversion (GGI).} Zhu et al.~\citep{Zhu_Huang_Xie_2024b} introduced GGI to address the limitations of traditional gradient inversion methods, which rely on gradient matching to reconstruct virtual data but are often bogged down by high-dimensional search spaces. Instead of directly optimising dummy images, GGI leverages low-dimensional latent vectors generated through a pre-trained generator, significantly reducing the complexity of the search process. Compared with existing methods, this approach not only improves the efficiency of image reconstruction, but also enhances the quality of restored images. 

\textbf{Gradient Inversion via Neural Architecture Search (GI-NAS).} Attributed to Yu et al. ~\citep{Yu_Fang_Chen_Sui_Chen_Wu_Xia_Xu_2024}, GI-NAS overcomes the shortfall of existing gradient reversal approaches, which rely on explicit prior knowledge, such as pre-trained generative models. Unlike traditional methods that use fixed architectures, GI-NAS adaptively searches for neural architectures to exploit implicit prior knowledge, enhancing the adaptability and generalization of attacks in various settings. Extensive experiments showed that GI-NAS outperforms many existing methods, especially under challenging conditions such as high-resolution images, large batch sizes, and advanced defense mechanisms.

\subsection{Generative Model-based Attacks}

With the emergence of Generative Adversarial Networks (GANs) as highly effective generative frameworks, recent MI attacks exploit the generative capabilities of GANs to reconstruct sensitive input data from the outputs or intermediate features of a trained machine learning model, even in black-box scenarios. In these attacks, adversaries leverage the learned representations of a model to recover private information, such as facial images, by generating high-quality approximations of the original data. By employing an adversarial process—where a generator network endeavors to recreate input samples that align with the model’s internal representations and a discriminator network assesses the realism of these reconstructions—GAN-based MI attacks can effectively destroy privacy protection, thereby posing significant risks to data confidentiality~\citep{Pang_Chen_Deng_Wu_Bai_Xu_2024}.

\textbf{Generative Model Inversion (GMI) and its Improved Version Variational Model Inversion(VMI).} Zhang et al.~\citep{Zhang_Jia_Pei_Wang_Li_Song_2020} presented GMI that effectively utilizes DNNs to reconstruct sensitive training data. By leveraging GANs to learn informative priors from public datasets, the GMI attack is an improvement over previous model inversion methods, which struggle with complex architectures and high-dimensional data spaces. GMI substantially increases reconstruction quality, especially in tasks like face recognition, where it achieves high recognition accuracy. The key innovation is the use of a generative model to guide the inversion process, which offers a more realistic and semantically rich reconstruction of privacy data than earlier approaches such as basic MI attacks that rely on optimisation techniques. Wang et al.~\citep{Wang_Fu_Li_Khisti_Zemel_Makhzani_2021} introduced VMI, a variational framework that formulates the MI attack as a variational inference~(VI) problem. In this framework, a generative model (usually a GAN) is trained on an auxiliary dataset so that the attack can efficiently learn a prior distribution similar to the target data. This approach is superior to previous MI techniques, especially in terms of restoring more realistic and diverse samples while retaining a high level of target accuracy. By optimising a variational target containing a Kullback-Leibler divergence term, the attack allows for control over the trade-off between realism and accuracy, resulting in a more flexible and efficient reconstruction of sensitive data, such as faces or medical images. The VMI approach is built on GAN-based MI techniques (e.g., Zhang et al.'s GMI attack~\citep{Zhang_Jia_Pei_Wang_Li_Song_2020}), but expands the functionality of these techniques by integrating probabilistic models and exploiting more meaningful normalised streams in the latent space.

\textbf{KEDMI.} Proposed by Chen et al.~\citep{Chen_Kahla_Jia_Qi_2021}, KEDMI is a novel GAN optimised for specific inversion tasks. 
The core of KEDMI is training a discriminator that not only differentiates between real and generated samples, but also discriminates the soft labels provided by the target model. This approach enables the generator to retain image statistics that are more relevant to the class inference of the target model, which are also most likely to be present in the private training data that is unknown. In addition, the authors introduced a distribution recovery method that does not recover individual data points, but rather to restore the distribution of training samples corresponding to a given label. Such a many-to-one assumption is more realistic, since a classifier naturally correlates to multiple training samples. The experiments showed that this knowledge-rich distribution inversion attack method significantly improves the success of the attack and demonstrates its validity across a wide variety of datasets and network architectures. This work strengthens the performance of MI attacks and offers new ideas about improving GANs for specific attack tasks.

\textbf{Pseudo Label-Guided MI (PLG-MI).} Yuan et al.~\citep{Yuan_Chen_Zhang_Zhang_Yu_Zhang_2023} designed the PLG-MI attack, which uses a conditional GAN (cGAN) to enhance the effectiveness of the MI attack by increasing visual quality and the success of reconstructed data. A notable novelty of PLG-MI is the use of a top-n selection scheme to generate pseudo-labels from public data, decoupling the way the cGAN is trained from different classes in the search space, and reducing the scope of the optimisation problem. In addition, the authors introduced the max-margin loss, which further improves the search process by focusing on the subspace of specific target categories. Extensive experiments showed that PLG-MI performs better than previous GAN-based attacks with significant changes in distribution, enabling higher-fidelity reconstruction and improved attack success across a variety of datasets and models.

\textbf{Intermediate Features Enhanced Generative Model Inversion (IF-GMI).} Developed by Qiu et al.~\citep{Qiu_Fang_Yu_Chen_Qiu_Xia_2024}, IF-GMI seeks to enhance the performance of MI attacks by utilising intermediate features of the GAN architecture, which improves over existing MI attack approaches that depend only on latent space representations in GANs. 
By breaking down the GAN architecture and not only optimising the latent code but also the intermediate features, IF-GMI significantly increases the representational power of the model, resulting in a more realistic and semantically accurate reconstruction of private data. The authors implemented L1 ball constraints during the optimisation to prevent the generation of unrealistic images, which allowed the method to outperform existing MI attacks on multiple benchmarks, especially in out-of-distribution (OOD) scenarios.

\textbf{MIRROR.} An et al.~\citep{An_Tao_Xu_Liu_Shen_Yao_Xu_Zhang_2022} proposed MIRROR, a leading-edge methodology in the area of GAN-based MI attacks, demonstrating remarkable progress in the fidelity of reversed samples. This work is based on the distinctive architecture of StyleGAN, which helps to break down input styles at different granularities. This enables the model to learn and regenerate these styles individually during training, a process that is especially beneficial for MI attacks. Given the target labels, MIRROR takes a StyleGAN trained on public data and applies either gradient descent or genetic search algorithms together with distribution-based clipping to locate the optimal parameterisation of the styles. The samples thus generated not only conform to the target labelling criteria set by the subject model, but are also recognizable by humans. It is shown that MIRROR's reversed samples exhibit higher fidelity than existing techniques, advancing MI attacks in the privacy threat domain.

\textbf{Plug \& Play Attacks (PPA).} Early studies (e.g., GMI~\citep{Zhang_Jia_Pei_Wang_Li_Song_2020}, VMI~\citep{Wang_Fu_Li_Khisti_Zemel_Makhzani_2021}) demonstrated the utility of GANs as image priors tailored to specific models, but these methods are found to be resource-intensive, inflexible, and vulnerable to dataset distributional shifts. Struppek et al.~\citep{Struppek_Hintersdorf_Correia_Adler_Kersting_2022} devised PPA, an innovative approach that decouples the dependency between the target model and the image prior, allowing for the use of a single pre-trained GAN to attack multiple targets with minimal adjustments. This study revealed that these attacks remain robust under strong distributional shifts, while producing high-quality images that expose sensitive class features.

\textbf{Boundary-Repelling Model Inversion (BREP-MI).} Kahla et al.~\citep{Kahla_Chen_Just_Jia_2022} proposed the BREP-MI algorithm, demonstrating the viability of MI using only the labelling information of model predictions. In this context, the attacker only has access to the output labels of the model and not any confidence scores or soft labels, thus making it much more challenging. The key novelty of BREP-MI is its boundary exclusion mechanism, whereby the predictive labels of the model are assessed in a spherical region, guiding the attacker to the direction of the centroid of the estimated target category. The BREP-MI approach proved to be capable of high-fidelity reconstruction of private training data (e.g., face images) even in the absence of detailed model knowledge. The authors compared BREP-MI with existing white-box and black-box MI attacks and found that BREP-MI performs better than black-box methods and is close to the effectiveness of white-box attacks, but requires much less model information.

\textbf{Coarse-to-Fine Model Inversion (C2FMI) Attacks.} As the scope of the threat of MI attacks expands, black-box scenarios in which attackers have limited access to models have received much attention. Ye et al.~\citep{Ye_Luo_Naseem_Yang_Shi_Jia_2023} introduced C2FMI, a two-phase attack aimed at increasing the effectiveness of MI attacks under black-box situations. The first phase of C2FMI involves an inversion network that ensures that the reconstructed image (known as the attack image) is located near the stream shape of the training data. In the second phase, the attack uses a black-box oriented tactic to refine these images so that they closely resemble the original training data. Notably, C2FMI even outperforms some white-box MI attacks, which manifests the validity of its two-phase design. This work also gave a robustness analysis on assessing the consistency of C2FMI in comparison with existing MI attacks, offering insights into the robustness of the attack. Moreover, the authors explored the potential defenses against C2FMI.

\textbf{RLBMI.} Han et al.~\citep{Han_Choi_Lee_Kim_2023} presented a Reinforcement Learning (RL)-based method for black-box MI attacks to address the limitations of GANs. The authors designed the search for private data in the latent space as a Markov decision process, using RL to guide the exploration of this space and rewarding the confidence score of the generated images. This new method enables effective navigation through the potential space and greatly increases attack performance compared to existing black-box GAN-based attacks. Experiments performed on a variety of datasets and models demonstrate that this approach delivers strong performance, and that it successfully reconstructs private training data in black-box scenarios with fewer queries and higher accuracy than traditional black-box-based GAN methods.

\textbf{LOKT.} Nguyen et al.~\citep{Nguyen_Chandrasegaran_Abdollahzadeh_Cheung_2024} investigated the scenario of a challenging label-only MI attack, where the adversary only gets access to the predictive labels of the model with no additional confidence scores or model information. To tackle the issue, the authors proposed a promising technique named LOKT, which facilitates the knowledge transfer from the opaque target model to the agent model. This approach exploits generative modelling techniques, particularly through using a target model-assisted conditional generative adversarial network, to achieve effective knowledge transfer and advanced white-box MI attacks in a previously restricted setting. The experimental results demonstrated that LOKT beats existing label-only MI attacks by over 15\% on various benchmarks, while providing optimised query budgets. This research highlights that even the least important information (e.g., hard labels) can be exploited to reconstruct sensitive training data.

\textbf{SecretGen.} Yuan et al.~\citep{Yuan_Wu_Long_Xiao_Li_2022} developed SecretGen, a novel approach to recovering private information from pre-trained models using the generative power of GANs. The work is particularly relevant to GAN-based MI attacks as it solves the problem of extracting sensitive data in the absence of ground truth tags. SecretGen is composed of a conditional GAN that acts as a generative backbone, a pseudo-tag predictor, and a latent vector selector. These components work together to produce realistic images that may resemble private training data for the target model. SecretGen works efficiently in both white- and black-box settings, proving its usefulness in real-world scenarios. The extensive experiments showed that SecretGen can achieve performance comparable to or better than existing methods, even if the latter can access ground truth labels. In addition, the authors provided a comprehensive analysis of SecretGen's performance and its resistance to sanitised defenses, shedding light on the importance of privacy-preserving practices in the era of transfer learning.

\textbf{Gradient Inversion over Feature Domains (GIFD).} Fang et al.~\citep{Fang_Chen_Wang_Wang_Xia_2023} presented GIFD  to augment privacy attacks in FL systems. In contrast to traditional gradient inversion attacks that rely on pre-trained GANs running in the latent space, limiting their expressiveness and generality, GIFD operates by breaking down the GAN model and performing optimisations in the feature domain in the middle layer. This layered approach enables more accurate gradient inversion, refining the optimisation progressively from the latent space to feature layers that are closer to the output image. Additionally, the authors proposed an l1-sphere constraint to mitigate the risk of generating unrealistic images, and extended the method to an OOD setting, which addresses the mismatch between GANs' training data and the FL task. The experiments showed that GIFD attains superior pixel-level reconstruction, while being effective under various defense strategies and batch sizes, rendering it a robust tool for evaluating privacy vulnerabilities in FL systems.

\textbf{Dynamic Memory Model Inversion Attacks (DMMIA).} Built on the concept of leveraging historical knowledge, Qi et al.~\citep{Qi_Chen_Mao_Hui_Li_Zhang_Xue_2023} designed DMMIA to enhance attack performance. DMMIA employs two prototype-based representations: intraclass multicentre representation~(IMR) and interclass discriminative representation~(IDR). The IMR captures target-related concepts through multiple learnable prototypes, and the IDR describes memory samples as prototypes to extend privacy-relevant information. The use of these prototypes allows DMMIA to generate more varied and discriminative outcomes than existing MI attack methods, as evidenced by DMMIA's excellent performance on various benchmark datasets.

\textbf{Patch-MI.} Inspired by puzzle assembly, Jang et al.~\citep{Jang_Lyu_Yang_2023} proposed Patch-MI to reinforce MI attacks through patch-based reconstruction techniques. Traditional generative MI attacks typically depend on auxiliary datasets, and the success depends on the similarity among these datasets and the target dataset. When the distributions are different, these methods often yield impractical reconstructions. To resolve the issue, Patch-MI adopts a patch-based discriminator in a GAN-like framework to reconstruct images even when auxiliary datasets are dissimilar. Patch-MI also incorporates random transform blocks to increase the ability to generalise reconstructed images, which eventually improves the accuracy of the target classifiers. Patch-MI outperforms existing MI methods while preserving comparable statistical quality, marking a big step forward in the field of MI attacks.

\textbf{SIA-GAN.} Challenging the presumed security of scrambling methods, Madono et al.~\citep{Madono_Tanaka_Onishi_Ogawa_2021} proposed the SIA-GAN model by mapping scrambled images back to their original form through learning. The findings of the work indicate that certain transformations (e.g., block shuffling) hinder reconstruction, but the security of scrambling techniques has not been adequately investigated.

\textbf{FedInverse: A GAN-based Framework for MI Attack Assessment.} Wu et al.~\citep{Wu_Bai_Song_Chen_Zhou_Xiang_Sajjanhar_2024} presented FedInverse to evaluate the risk of MI attacks in FL systems, in which an adversary who masquerades as a benign participant can use a shared global model to reconstruct other participants' privacy data. In spite of continuous advances in defense mechanisms, FedInverse reveals that current defenses are still ineffective against sophisticated MI attackers. The authors optimised their approach using the Hilbert-Schmidt independence criterion to increase the diversity of MI attacks generated. FedInverse is tested with three types of MI attacks (i.e., GMI, KEDMI, and VMI) and demonstrated its effectiveness in reconstructing private data from FL participants. It highlights the necessity of assessing the risk of privacy breaches in FL.

\subsection{Optimisation-based Attacks} \label{optim}
In MI attacks, optimisation-based methods refine sensitive data reconstruction by iteratively minimizing loss functions so that the models' outputs or internal representations align with specific criteria. These methods are highly versatile and suitable for a variety of scenarios, such as white-box attacks (where model gradients are accessible) and black-box attacks (where only predictions are used). The reconstruction is guided by well-designed loss functions, including gradient matching loss (minimizing the difference between the observed gradient and the gradient computed from the reconstructed inputs) and output matching loss (minimizing the discrepancy between the reconstructed inputs and the model outputs from the target inputs)~\citep{Guo_Zeng_Chen_Zhang_Ren_Zhou_Qu_2024}. Utilizing an optimisation algorithm such as gradient descent or stochastic gradient descent, these methods iteratively adjust the reconstructed inputs to make high-fidelity data recovery~\citep{Zhu_Liu_Han_2019, Fang_Qiu_Yu_Yu_Kong_Chong_Chen_Wang_Xia_Xu_2024, Guo_Zeng_Chen_Zhang_Ren_Zhou_Qu_2024}.

\textbf{Gradient Matching}: An attacker can utilize the gradient updates shared during the FL process to reconstruct individual training data. To be specific, the attacker minimizes the difference between the observed gradient and the gradient computed from the reconstructed input. Most of the attacks in the gradient inversion category, such as DLG~\citep{Zhu_Liu_Han_2019}, iDLG~\citep{Zhao_Mopuri_Bilen_2020}, and LOMMA~\citep{Nguyen_Chandrasegaran_Abdollahzadeh_Cheung_2023}, fall in this catergory.

\textbf{Output Matching}: An attacker reconstructs inputs that generate outputs similar to the target model. The data is reconstructed through matching the confidence scores of the target model~\citep{Guo_Zeng_Chen_Zhang_Ren_Zhou_Qu_2024}. For instance, the study in \citep{Fredrikson_Lantz_Jha_Lin_Page_Ristenpart_2014} shows that adversarial access to linear classifiers exposes sensitive genomic information in personalized medicine. 
Based on this notion, the authors introduced a novel MI attack that utilizes confidence values output by machine learning models to make these attacks applicable to a variety of settings.


\subsection{Miscellaneous MI Attacks}
The following MI attacks do not fit into those well-defined categories in Sections~\ref{gradient_inv}-\ref{optim}, and are therefore included in this section. 

\textbf{DeepInversion.} Yin et al.~\citep{Yin_Molchanov_Alvarez_Li_Mallya_Hoiem_Jha_Kautz_2020} introduced DeepInversion to synthesize high-quality images from pre-trained neural networks that do not need access to the original training data. Unlike traditional MI techniques which may require auxiliary information or shallow models, DeepInversion uses internal feature statistics stored in the batch normalization layer to generate realistic images. DeepInversion is well suited for tasks such as data-free knowledge refinement, pruning, and continuous learning. DeepInversion's ability to synthesize images with high levels of fidelity and contextual accuracy sets it apart from earlier techniques such as the MI attacks of Fredrikson et al.~\citep{Fredrikson_Jha_Ristenpart_2015}, which focus on reconstructing base class images via gradient descent methods. While GANs are widely used for generative modeling, DeepInversion provides an efficient alternative, making it an important contributor to privacy-relevant inversion attacks and knowledge transfer scenarios.

\textbf{RL-GraphMI.} Zhang et al.~\citep{Zhang_Liu_Huang_Wang_Lee_Chen_2022} studied MI attacks against graph neural networks~(GNNs), highlighting the privacy risks inherent in graph data because of its relational structure. The authors emphasized that while MI attacks are effective in lattice-like domains, their application to non-lattice structures (e.g., graphs) tends to produce sub-optimal results due to the uniqueness of graphs. To tackle this problem, the authors introduced GraphMI, a novel approach that incorporates a projected gradient module to manage graph edge discretization and maintain feature sparsity, as well as a graph autoencoder to effectively exploit graph topology and attributes. Furthermore, the authors introduced RL-GraphMI for hard-labeled black-box settings, which uses gradient estimation and reinforcement learning to assess the risk of MI associated with edge effects. Through extensive experiments on public datasets, the authors validated RL-GraphMI, and also assessed two existing defense methods, differential private training and graph preprocessing, finding that they are not sufficient to defend against privacy attacks.

\textbf{Explainable Artificial Intelligence (XAI).} MI attacks pose substantial privacy risks, particularly when model interpretability is augmented with interpretable XAI. Zhao et al.~\citep{Zhao_Zhang_Xiao_Lim_2021} investigated these risks and identified a wide variety of attack architectures that leverage model interpretation for the reconstruction of private image data, proving that such interpretation may unintentionally expose sensitive information. The authors developed a multimodal transposition CNN architecture that exhibits a significantly higher inversion performance than methods that rely solely on target model predictions. The results indicate that the spatial knowledge embedded in image interpretation can be used to improve attack efficiency. Moreover, the authors emphasized that even with an unexplained target model, inversion performance can be improved by an attention-shifting approach that reconstructs the target image without the target model by first inverting the interpretation from the proxy model. This research shows that while XAI helps user understanding, it increases vulnerability to privacy attacks, and thus new privacy-preserving techniques are required to balance the dual requirements of AI interpretability and privacy.

\textbf{Supervised Model Inversion (SMI).} Tian et al.~\citep{Tian_Cui_Zhang_Tan_Yu_Tian_2023} explored the usefulness of class information in MI attacks and proposed SMI. This approach decreases the reliance on a priori target information by learning pixel-level and data-to-category features from the victim model's polled output and labeled auxiliary datasets. Their findings indicate that the inversion samples generated by SMI are visually more convincing and richer in detail than existing MI methods. In addition, this study found that the makeup of the auxiliary dataset plays a vital role in determining the quality of the reconstructed samples, and that ground truth labeling, while useful, is not essential for a successful attack.

\textbf{Label Smoothing Model Inversion (LSMI).} Struppek et al.~\citep{Struppek_Hintersdorf_Kersting_2023} studied the effects of label smoothing, a common regularization technique, on model vulnerability to MI attacks. The findings suggest that traditional label smoothing softens class labels to enhance generalization and calibration, but inadvertently increases the model's vulnerability to MI attacks, thereby magnifying privacy leakage. To address the issue, the authors introduced LSMI, a new label smoothing approach through applying a negative factor. LSMI shows greater resilience against MI attacks and hinders the extraction of category-relevant information, outperforming existing MI defense methods.

\textbf{Ensemble Model Inversion (EMI).} Wang and Kurz~\citep{Wang_Kurz_2022} presented EMI, which enhances traditional MI methods by simultaneously utilizing multiple trained models to infer the distribution of the original training data. Using ensemble models provides the adversary with a much richer perspective, resulting in a higher quality reconstruction of training samples. EMI significantly improves single-model inversion, allowing the generated samples to display distinguishable features of dataset entities. Furthermore, the authors explored the effectiveness of EMI in the absence of reliance on auxiliary datasets, showing high-quality data-free MI. This study shows the importance of model diversity in ensembles and includes additional constraints to boost prediction accuracy and reconstructed sample activation.

\subsection{Summary}

Gradient inversion attacks, generative model-based attacks, and optimisation-based attacks are at the core of MI. Each of them has different methods and applications. 

Gradient inversion attacks utilize the gradient information shared during federated or distributed training to reconstruct sensitive data. Methods such as DLG and iDLG allow for the alignment of fake gradients with real ones, while more advanced methods like AGIC improve scalability and accuracy. These attacks are efficient in either white- or gray-box settings, but rely on gradient access, limiting their scope. 

Generative model-based attacks take advantage of frameworks such as GAN to reconstruct data by approximating data distributions guided by model outputs or intermediate features. Techniques such as GMI and VMI improve fidelity by leveraging latent spatial priors or probabilistic modeling, and are thus effective in both white- and black-box settings. Nevertheless, they typically need auxiliary datasets or pre-trained models, which adds complexity and resource requirements. 

Optimisation-based attacks, covering aspects of gradient inversion attacks and generative model-based attacks, iteratively refine the reconstructed data by keeping inputs consistent with outputs, gradients, or intermediate features via tailored loss functions. While these attacks have the accuracy of gradient inversion and the flexibility of generative methods, they are  computationally resource-intensive and rely on specific attack inputs, highlighting the challenges of defending against various MI threats. 

\begin{figure}[h!]
  \centering
  \includegraphics[width=0.9\textwidth]{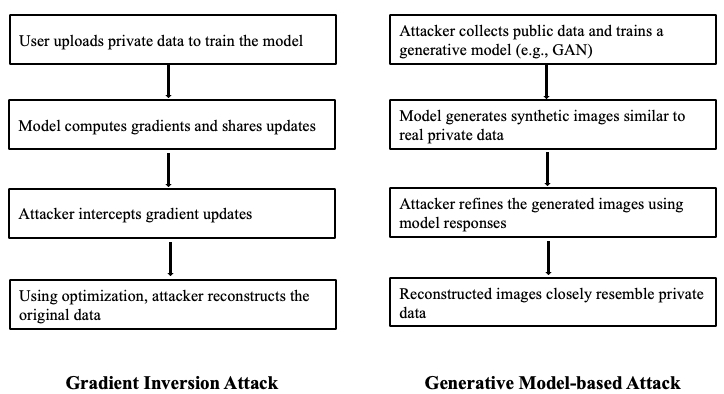}
  \caption{A demonstrative example of gradient inversion attacks vs. generative model-based attacks.}
  \label{fig_GIA_VS_GMA}
\end{figure}

Figure~\ref{fig_GIA_VS_GMA} presents an illustrative example of gradient inversion attacks vs. generative model-based attacks. Moreover, structured around a host of key factors, Table~\ref{table_gradien_vs_gan} presents a detailed comparison of gradient inversion attacks and generative model-based attacks. 

\begin{table}[h!]
\centering
\renewcommand{\arraystretch}{1.5} 
\setlength{\tabcolsep}{6pt} 
\begin{tabular}{|p{2.5cm}|p{4.5cm}|p{5cm}|}
\hline
\textbf{Key Factors}                   & \textbf{Gradient Inversion Attacks}                          & \textbf{Generative Model-based Attacks}                \\ \hline
\textbf{Primary Methodology}       & Gradient matching through optimisation                 & Leverages generative models (e.g., GANs)         \\ \hline
\textbf{Input Dependency}          & Requires gradient updates                             & Uses outputs or intermediate representations      \\ \hline
\textbf{Flexibility}               & Limited to scenarios with gradient access             & Applicable in both white-box and black-box settings \\ \hline
\textbf{Computational Complexity}  & Relatively low due to direct gradient optimisation     & Higher, due to generative modeling and training   \\ \hline
\textbf{Auxiliary Data Requirement} & Often unnecessary                                      & Frequently relies on auxiliary datasets           \\ \hline
\textbf{Reconstruction Fidelity}   & High for gradient-specific data                       & High, particularly for complex distributions      \\ \hline
\textbf{Application Scenarios}     & Federated learning or gradient-sharing settings       & General machine learning settings                \\ \hline
\end{tabular}
\caption{Comparison of gradient inversion attacks and generative model-based attacks.}
\label{table_gradien_vs_gan}
\end{table}

\subsection{MI in Different Applications}
\label{sec:Different Applications}
MI attacks pose a significant privacy threat across various domains where deep learning models process sensitive data. These attacks exploit learned representations to reconstruct private inputs, resulting in privacy and data breaches in biometric recognition systems, healthcare, and financial services.

\subsubsection{MI in Biometric Recognition Systems}
Biometric systems, especially those used for facial recognition~\citep{Zhang_Hou_Yang_Yang_Yang_Cui_2024}, fingerprint authentication, and iris scanning, rely heavily on deep learning models to process and match individuals' unique biological traits. These systems are often deployed in high-stakes environments such as border control, law enforcement, and secure access control. MI attacks on these systems can result in the reconstruction of highly sensitive data (e.g., an individual's facial image or fingerprint). This poses severe privacy risks, as biometric data, once compromised, cannot be easily changed or replaced like passwords or other traditional security measures~\citep{Yang_Wang_Hu_Ibrahim_Zheng_Macedo_Johnstone_Valli_2019, Yang_Wang_Hu_Zheng_Valli_2019, Yang_Wang_Kang_Johnstone_Bedari_2022}. By reconstructing biometric data, adversaries can bypass authentication systems, impersonate individuals, or misuse this data for malicious purposes. Thus, studying MI attacks and defenses~\citep{Ahmad_Mahmood_Fuller_2022, Huang_Wang_Li_Yang_Song_Wang_2024, Khosravy_Nakamura_Hirose_Nitta_Babaguchi_2022, Khosravy_Nakamura_Hirose_Nitta_Babaguchi_2021, Yoshimura_Nakamura_Nitta_Babaguchi_2021} is critical to ensuring the security and privacy protection of biometric systems.

\subsubsection{MI in Healthcare Systems}
Healthcare is another area where deep learning models are widely employed to aid in diagnosing diseases, predicting patient prognosis, and planning treatments~\citep{Ra_Li_Li_2021, Pei_Li_Siuly_Wen_2022, Wei_Li_Yang_2023}. These models often deal with highly sensitive medical records, including patient history, imaging data (e.g., X-rays or MRIs)~\citep{Dao_Nguyen_2024}, and genetic data. MI attacks on healthcare models present a serious threat because it is likely that  adversaries can recover private health information from the model's output. Exposure of healthcare data through MI attacks can lead to many forms of harm, including identity theft, insurance fraud, and discrimination due to medical conditions ~\citep{Hatamizadeh_Yin_Molchanov_Myronenko_Li_Dogra_Feng_Flores_Kautz_Xu_2023}. Furthermore, improper handling of such sensitive data may have severe legal consequences for healthcare providers, undermine patient trust, and jeopardize the credibility of AI-based healthcare solutions. Understanding and mitigating MI attacks in this context is therefore vital to patient privacy protection and ethical deployment of AI in healthcare.

\subsubsection{MI in Financial Systems}
In finance, deep learning models are increasingly used for tasks such as fraud detection, credit scoring and risk evaluation~\citep{Milner_2024}. These models run on large amounts of sensitive financial data, including transaction history, personal financial information, and credit history. MI attacks on financial models allow adversaries to infer private financial details, reconstruct specific transactions, resulting in financial fraud or identity theft, or other severe financial consequences for individuals and organizations. Personal financial data, once exposed, can be used for unauthorized transactions or to compromise other systems. In addition, successful MI attacks could damage confidence in AI-based financial services that rely on trust and confidentiality~\citep{Galloway_Karakolios_Ma_Perdisci_Keromytis_Antonakakis_2024}. Thus, it is important for financial institutions to take the potential impact of MI attacks seriously and develop defensive measures to ensure the integrity of financial models and the security of customer data.

\subsection{MI on Different Data Types}
\label{sec:Different Data Types}

\begin{figure}[h!]
  \centering
  \includegraphics[width=0.9\textwidth]{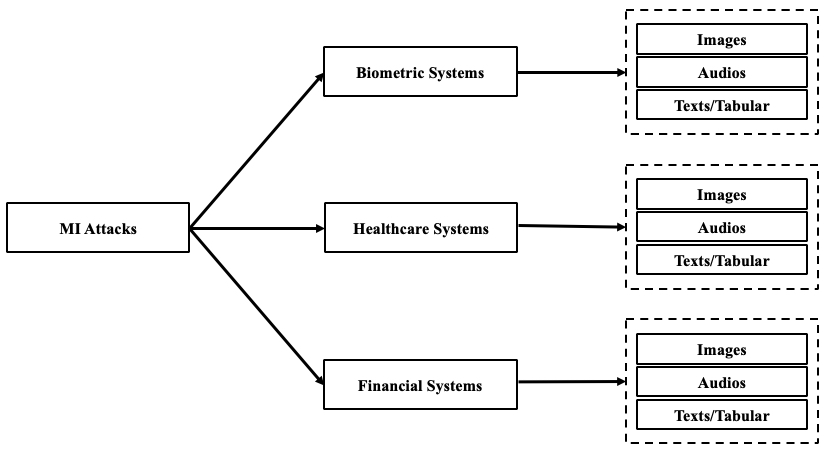}
  \caption{MI attacks across various applications where different data types are processed.}
  \label{Fig_MI_Applications_Data_Types}
\end{figure}

MI attacks across different application domains, including biometric systems, healthcare systems, and financial systems and each of these applications processes different data types, such as images, audio, and text/tabular data, which are susceptible to MI attacks, as demonstrated in Figure~\ref{Fig_MI_Applications_Data_Types}. The latest advances in MI attacks increase the vulnerability of these data types to MI attacks. MI attacks leverage the learned parameters, features or outputs of deep learning models to reconstruct sensitive training data, thereby compromising the privacy of user information. As deep learning systems increasingly penetrate areas such as visual recognition, speech processing, natural language understanding, and analysis of structured data, the possibility of privacy breaches by malicious attacks is increasing.

\subsubsection{MI on Images}
MI attacks expose a critical privacy vulnerability in deep learning systems, particularly in applications dealing with image data. These attacks aim to reconstruct input images from the outputs or parameters of trained models, compromising the confidentiality of training data. In image-based systems, MI attacks can exploit gradient information~\citep{Geiping_Bauermeister_Dröge_Moeller_2020, Zhu_Liu_Han_2019, Zhao_Mopuri_Bilen_2020}, or model outputs (e.g., confidence scores~\citep{Guo_Zeng_Chen_Zhang_Ren_Zhou_Qu_2024}) to recover visual details of training samples, including personally identifiable information. Addressing these risks has led to the development of defensive strategies, such as feature perturbation~\citep{Sun_Li_Wang_Yang_Li_Chen_2021}, cryptographic techniques~\citep{Prakash_Ding_Li_Errapotu_Pei_Pan_2020} and differential privacy~\citep{Li_Yu_Cheng_Yan_Zhang_2024}, against the MI attacks targeting images.

\subsubsection{MI on Audios}
MI attacks on audio data create distinct and pressing challenges for privacy protection in machine learning systems. These attacks exploit model parameters or outputs to reconstruct sensitive audio inputs, such as spoken phrases or unique acoustic features, potentially exposing private or proprietary information. In view of the inherently rich temporal and spectral characteristics of audio data, MI attacks can disclose subtle features of the training samples, including speaker identity or confidential speech content. For example, Pizzi et al.~\citep{Pizzi_Boenisch_Sahin_Böttinger_2023} extended the application of MI attacks to automatic speech recognition systems, uncovering the potential for voice extraction from trained models. The authors introduced a new technique called sliding model inversion, which augments traditional MI attacks by iteratively inverting overlapping audio chunks. This approach makes effective use of the sequential nature of audio data, accomplishing the reconstruction of audio samples and the extraction of intermediate speech features, thus offering important lessons for speaker biometrics. The authors demonstrated that inverted audio can be used to generate spoofed samples to impersonate a loudspeaker and execute voice-protected commands in a secure system.

\subsubsection{MI on Texts and Tabular Data}

MI attacks on texts and tabular data reveal privacy vulnerabilities in natural language processing~(NLP) models, in which sensitive information such as personally identifiable information, confidential communications, or proprietary content can be reconstructed from model outputs or learned representations. These attacks use patterns in embedding vectors~\citep{Petrov_Dimitrov_Baader_Müller_Vechev_2024}, labeling probabilities, or gradients to infer training data, creating risks for applications that range from sentiment analysis to language generation. The structural and context-dependent nature of textual data amplifies the likelihood of sensitive reconstruction, especially if the training dataset contains private or domain-specific information. MI-related studies on texts and tabular data are reviewed below.

\textbf{DAGER.} Petrov et al.~\citep{Petrov_Dimitrov_Baader_Müller_Vechev_2024} proposed the DAGER algorithm, designed specifically for accurate gradient inversion of large language models. DAGER addresses a major drawback of earlier research that focused on image data. While initial attacks on the text domain were restricted to approximate reconstruction of small or short input sequences, DAGER overcomes these inadequacies by utilizing the low-rank structure of the gradient of the self-attentive layer and the discrete nature of the token embedding vectors. The authors showed that DAGER can achieve accurate recovery of entire batches of input texts with the use of efficient GPUs, outperforming existing approaches in terms of speed, scalability, and reconstruction quality.

\textbf{Text Revealer.}  Text Revealer~\citep{Zhang_Hidano_Koushanfar_2022} is the first systematic study of MI attacks for text reconstruction in transformer-based classifiers. The study showed how to faithfully reconstruct private training data by utilizing external datasets and GPT-2 to produce class-domain text and then perturb it based on feedback from the target model. A large number of experiments were conducted to demonstrate the effectiveness of their attack on text datasets of varying lengths and the successful reconstruction of sensitive text information.

\textbf{TabLeak.} Proposed by Vero et al.~\citep{Vero_Balunović_Dimitrov_Vechev_2023}, TabLeak is the first comprehensive MI attack tailored to tabular data. TabLeak is able to achieve over 90\% reconstruction accuracy on private data despite large batch sizes, thus destroying assumptions about the security of FL protocols such as FedSGD and FedAvg. Earlier research in other domains (e.g., images) has found similar threats. Nevertheless, the distinctive challenges of dealing with mixed types of tabular data with discrete and continuous features require new solutions such as soft-max relaxation and entropy-based uncertainty quantization.

\subsection{MI on Emerging Generative AI and Foundation Models}

With the recent rise of generative AI and foundation models like diffusion models (e.g., Stable Diffusion, DALL-E) and Large Language Models (LLMs) such as GPT-4, increased concerns about MI attacks have arisen. In contrast to traditional deep learning architectures, these models produce highly complex outputs and are therefore particularly vulnerable to data reconstruction attacks~\citep{Feretzakis_Papaspyridis_Gkoulalas_Divanis_Verykios_2024}. Attacks, such as MI attacks, seek to reconstruct training data from model outputs, raising ethical and security concerns. MI attacks leverage the model's learnt representations to restore sensitive training data, posing substantial privacy risks in applications involving text, images, and audios. It is worth noting that MI attacks present significant privacy and security threats to generative AI models, including diffusion models, LLMs, and other foundation models. Although privacy-preserving techniques continue to advance, many challenges remain.

\subsubsection{MI on LLMs}
As deep learning evolves, LLMs (e.g., OpenAI's GPT, Anthropic's Claude, and Meta's Llama) have greatly enhanced the ability to efficiently process a variety of downstream NLP tasks and unify them into a generative pipeline. Nevertheless, unrestricted access to these models also brings about potentially malicious attacks, such as MI attacks~\citep{Li_Chen_Luo_Wang_Peng_Kang_Zhang_Hu_Chan_Xu_2024}.

Embedding techniques are able to transform textual data into rich, dense numerical representations that capture semantic and syntactic properties, and thus have become a cornerstone of LLM functionality~\citep{Liu_Yao_Wu_Qin_Lin_Ren_Chen_2024}.
Embedding privacy is critical because these embeds often contain sensitive information about user data. Embedding inversion attacks are a specific subset of MI attacks that focus on restoring the original input data ~\citep{Li_Chen_Luo_Wang_Peng_Kang_Zhang_Hu_Chan_Xu_2024} from feature embeddings. For example, Pan et al. ~\citep{Pan_Zhang_Ji_Yang_2020} transforms a text sequence into a set of words to perform multi-label classification to predict multiple words for given embeddings. Morris et al.~\citep{Morris_Zhao_Chiu_Shmatikov_Rush_2023} considered the issue of language model inversion, showing that the next-token probabilities contain a large amount of information about the preceding text. 
The approach demonstrates for the first time that language model predictions are mostly reversible. In a number of cases, the authors are capable of recovering inputs that are similar to the original text, sometimes even recovering the input text completely. 

Chen et al.~\citep{Chen_Lent_Bjerva_2024} studied the security of LLMs from the perspective of multilingual embedding inversion. Specifically, the authors defined the problem of black-box multilingual and cross-lingual inversion attacks with a special focus on cross-domain scenarios. The results show that multilingual models are more vulnerable to inversion attacks than monolingual models. Dai et al.~\citep{Dai_Lu_Zhou_2025} discovered a realistic attack surface of LLMs: privacy leakage of training data in decentralized training, and proposed the first activation inversion attack (AIA). The AIA utilizes a public dataset to construct a shadow dataset consisting of text labels and corresponding activations. Using this dataset, an attack model can be trained so as to reconstruct the training data from the activations in decentralized training. The authors conducted experiments on various LLMs and publicly available datasets to show the vulnerability of decentralized training to AIA. 

Shu et al.~\citep{Shu_Li_Dong_Meng_Zhu_2025} identified MI attacks in a split learning framework for LLMs, highlighting the necessity of security defenses. They introduced mutual information entropy for the first time to understand the information propagation of transformer-based LLMs and evaluated the privacy attack performance of LLM blocks. 
The authors proposed a two-stage attack system where the first stage projects representations into the embedding space, and the second stage uses a generative model to recover text from the embeddings. This work highlights comprehensively the potential privacy risks when deploying personalized LLMs at the edge side.
To reduce embedding inversion attacks, Liu et al.~\citep{Liu_Yao_Wu_Qin_Lin_Ren_Chen_2024} proposed Embedding Guard (Eguard). Eguard employs transformer-based projection networks and textual mutual information optimization to protect embeddings while retaining the utility of LLMs. This approach greatly reduces privacy risks and protects more than 95\% of the tokens from being inverted, while maintaining high performance consistent with the original embeddings in downstream tasks.

\subsubsection{MI on Diffusion Models}

Diffusion models are emerging as favorable models for generating exceptionally high resolution image data. Diffusion models generate higher quality samples and are much easier to scale and control than previous models such as GAN. As a result, they have rapidly become the de facto approach for producing high-resolution images, and large-scale models such as DALL-E have attracted considerable public interest~\citep{Carlini_Hayes_Nasr_Jagielski_Sehwag_Tramer_Balle_Ippolito_Wallace_2023}.

However, diffusion models like stable diffusion are also subject to MI attacks that can extract training data from diffusion models. Chen et al.~\citep{Chen_Zhang_Bi_Hu_Hu_Xue_Yi_Liu_Tai_2025} provided a comprehensive review of recent advances in image inversion techniques, highlighting two main paradigms: GAN inversion and diffusion model inversion. In the context of diffusion model inversion, the authors explored training-free strategies, fine-tuning methods, and the design of additional trainable modules, and emphasized their unique advantages and limitations. Carlini et al.~\citep{Carlini_Hayes_Nasr_Jagielski_Sehwag_Tramer_Balle_Ippolito_Wallace_2023} demonstrated that diffusion models memorize individual images from training data and emit these images at generation time. By using a generate-and-filter pipeline, the authors pulled over a thousand training examples from state-of-the-art models, varying from photos of individual people to company trademark logos. They also trained hundreds of diffusion models in a variety of settings to analyze how different modeling and data decisions impact privacy. The findings suggest that diffusion models are much less private than previous generative models such as GANs.

Huang et al.~\citep{Huang_Hong_Chen_Roos_2024} investigated the privacy leakage risk of gradient inversion attacks. The authors designed a two-stage fusion optimization scheme that uses the trained generative model itself as prior knowledge to constrain the inversion search (latent) space, followed by pixel-level fine-tuning. The results show that the proposed optimization scheme can reconstruct an image almost identical to the original image. Wu et al.~\citep{Wu_Liu_Pu_Wei_Cao_Yao_2025} presented a practical gradient inversion method, namely Deep Generative Gradient Inversion, which utilizes the diffusion model's prior knowledge to improve the reconstruction performance for high-resolution datasets and larger batches. In addition, in order to address the spatial variation problem caused by the pre-trained diffusion model, group consistency regular terms are developed to constrain the distance between the reconstructed and aligned images. Zhang et al.~\citep{Zhang_Wei_Wu_Zhang_Zhang_Lei_Li_2024} proposed a means of directing the inversion process of a diffusion model toward a synthetic embedding of the core distribution. In addition, the authors introduced a spatial regularization method to balance the attention to the concepts being composed. This is devised as a post-training method that can be seamlessly integrated with other inversion approaches. Experimental results show that the proposed method is effective in mitigating the overfitting problem and generating more diverse and balanced combinations of concepts in synthetic images.

\section{Defenses Against MI Attacks}
\label{sec:Defenses}

This section outlines prominent and emerging defense strategies against MI attacks, divided into six categories: feature perturbation/obfuscation, gradient pruning, gradient perturbation/obfuscation, differential privacy, cryptographic encryption, and model/architecture enhancement. Each category has its unique approaches designed to protect data privacy while preserving model utility and computing efficiency.

\subsection{Feature Perturbation/Obfuscation} \label{perturbation}
Feature perturbation and obfuscation are crucial in alleviating MI attacks as they are designed to degrade the fidelity of the extracted data without undermining the utility of the underlying machine learning model. These techniques utilize various strategies, such as adding noise, transforming features, or hiding sensitive information, to distort data representations and thwart adversarial inference. Several feature perturbation and obfuscation approaches are summarized below.

\textbf{Vicinal Risk Minimization (VRM).} Ye et al.~\citep{Ye_Luo_Zhou_Zhu_Shi_Jia_2024} provided a thorough analysis of the success factors of GIAs and explained privacy risks inherent in distributed learning frameworks. The authors pointed out that while current defense strategies are abundant, they often degrade the performance of global models or require too much computational resources. This study also points to the gap in understanding the root cause of data leakage during distributed training. GIA is facing challenges in terms of model robustness, especially as changes in model structure can affect attack results. In response, the authors proposed, a plug-and-play defense scheme that leverages VRM and data augmentation based on neighborhood distributions, which as demonstrated, effectively enhances privacy without compromising model usability.

\textbf{Defense by Concealing Sensitive Samples (DCS2).} Wu et al.~\citep{Wu_Hayat_Zhou_Harandi_2024} put forward DCS2, a defense strategy against MI attacks in FL. FL is vulnerable to attacks that take advantage of low entanglement between the gradients, so the authors proposed a method to synthesize hidden samples. These samples imitate sensitive data at the gradient level while appearing different visually, thus obfuscating adversaries trying to reconstruct the data. Experimental results shown that DCS2 offers excellent protection for sensitive data without compromising the performance of FL systems, thereby setting a new standard for privacy-preserving defenses in FL.

\textbf{Automatic Transformation Search (ATS).} Gao et al.~\citep{Gao_Guo_Zhang_Qiu_Wen_Liu_2021} discussed and addressed a critical flaw in collaborative learning environments where the sharing of gradients can lead to the reconstruction of sensitive training data.  
The authors introduced ATS to identify carefully selected transformation policies that not only protect data privacy but also maintain the model's efficacy, obfuscating sensitive information. This work offers a practical solution to safeguard collaborative learning systems against gradient-based reconstruction attacks.

\textbf{Soteria.} Sun et al.~\citep{Sun_Li_Wang_Yang_Li_Chen_2021} proposed Soteria, a defense mechanism specifically designed to address the leakage of data representations within gradients, which is believed to be the main channel of privacy leakage in FL. The authors gave a comprehensive analysis of how data representations are embedded in model updates and described the potential for MI attacks. Soteria involves perturbing data representations to reduce the quality of reconstructed data while preserving the performance of the FL system. An important contribution of this work is the derivation of proven robustness and convergence guarantees for perturbing model updates, thereby ensuring the effectiveness of the defense without compromising accuracy. Empirical evaluations against attacks such as DLG using the MNIST and CIFAR10 datasets demonstrate that Soteria significantly improves privacy protection. This research provides valuable insights into the characterization of privacy breaches in FL and paves the way for more advanced defense strategies.

\textbf{Crafter.} Wang et al.~\citep{Wang_Ji_Xiang_Zhang_Wang_Zhou_Li_2024} proposed Crafter, a feature crafting mechanism to preserve identity information against adaptive MI attacks. Different to traditional methods based on adversarial games, Crafter misguides attackers by crafting features that direct them to non-privacy priors. These carefully crafted features can effectively function as poison training samples, restricting the attacker's capability to reconstruct private identities, even when adaptive counterattacks are employed. Experimental results show that Crafter performs better than existing countermeasures, successfully defending against both basic and adaptive MI attacks, while remaining functional for cloud-based deep learning tasks.

\textbf{Sparse-coding Architecture.} Dibbo et al.~\citep{Dibbo_Breuer_Moore_Teti_2024} explored a new approach to enhancing the robustness of neural networks against MI attacks, which utilize the output of the network to reconstruct private training data. The authors proposed an innovative network architecture that includes sparse coding layers, leveraging three decades of research on sparse coding in areas ranging from image denoising to adversarial classification. This work fills a gap in the understanding of how sparse coding can alleviate privacy vulnerabilities in neural networks. The experimental results demonstrate that sparse-coding architectures not only maintain a comparable or higher classification accuracy than existing defenses, but also reduce the quality of training data reconstruction across multiple metrics (PSNR, SSIM, FID) and datasets including CelebA, medical images, and CIFAR-10. In addition, the authors provided the cluster-ready PyTorch codebase to facilitate further research and standardization of defense evaluation against MI attacks.

\textbf{Synthetic Data Generation with Privacy-preserving Techniques.} Slokom et al.~\citep{Slokom_de_Wolf_Larson_2023} found that training deep learning models on unprotected raw data may lead to the disclosure of sensitive information, making individuals vulnerable to MI attribute inference attacks. To reduce this risk, the authors designed a two-step method that couples synthetic data generation with privacy-preserving techniques. Unlike previous methods that directly apply privacy measures to the original dataset, this approach first replaces the original data with synthetic data and then applies privacy-preserving techniques to maintain model performance, aiming to strike a balance between privacy preservation and model accuracy. The experiments show that this method decreases the success rate of MI attacks on “inclusive” individuals (present in the training data) and “exclusive” individuals (not included in the training data). 

\textbf{Image Augmentation.} Shin et al.~\citep{Shin_Boyapati_Suo_Kang_Son_2023} empirically analyzed the use of image enhancement as a defense strategy against MI attacks. The study shows that while differential privacy is a common solution, it often involves a trade-off of privacy versus utility. In comparison, image enhancement techniques are promising in alleviating such attacks without significantly impairing model performance. Through experiments on the CIFAR-10 and CIFAR-100 datasets, the authors identified the optimal combination of enhancements for different image classes, proving that these strategies work better than differential privacy. This study suggests that image enhancement may be used as a feasible alternative to minimize the loss of utility when defending against MI attacks.

\textbf{Additive Noise.} Titcombe et al.~\citep{Titcombe_Hall_Papadopoulos_Romanini_2021} indicated that even with limited information about data distribution, it is possible for a malicious computational server to successfully perform MI attacks that compromise user data privacy. In response, the authors proposed a simple noise addition method that can minimize the effectiveness of MI attacks while keeping an acceptable accuracy trade-off on the MNIST dataset. 

\textbf{Privacy-guided Training.} Goldsteen et al~\citep{Goldsteen_Ezov_Farkash_2020} presented a privacy-guided training method specifically designed for tree-based models. This approach lessens the impact of sensitive features in the model, thus decreasing the risk of feature-based inference without sacrificing the accuracy of the overall model. The authors remarked that sensitive features do not have to play a prominent role in model training, as privacy can be preserved by emphasizing less sensitive attributes. The authors showed the effectiveness of the proposed method in mitigating the risk of MI in both black-box and white-box scenarios.

\textbf{Statistical Features via Knowledge Distillation.} Gao et al.~\citep{Gao_Zhu_Ye_Zhou_2024} studied the vulnerability of FL to gradient inversion attacks, which allows the server to reconstruct the client's training data by reversing the uploaded gradients. To mitigate this risk, the authors proposed a defense mechanism that exploits statistical information/features derived from the training data rather than the data itself. Motivated by statistical machine learning, this method involves training lightweight local models via knowledge refinement so that the server can only access semantically meaningless statistics. The experiment results show that this method is superior to existing defense strategies in terms of reconstruction accuracy, model accuracy, and training efficiency.

\subsection{Gradient Pruning}

Gradient pruning, also known as gradient compression or sparsification, has emerged as a privacy-preserving machine learning technique against MI attacks. By either selectively pruning the gradient or reducing the gradient transmission, these methods are designed to enhance privacy preservation, improve communication efficiency, and preserve model performance. This section discusses gradient pruning methods for MI defense.

\textbf{PATROL.} Ding et al.~\citep{Ding_Zhang_Pan_Yuan_2024} proposed PATROL to deal with privacy risks in collaborative inference for edge devices. Although collaborative inference realizes advanced use of DNNs by unloading intermediate results to the cloud, it is susceptible to MI attacks. PATROL addresses this problem by reducing leakage of sensitive information through the use of privacy-oriented pruning, deployment of more layers at the edge, and focusing on task-relevant features. In addition, Lipschitz regularization~\citep{Virmaux_Scaman_2018} and adversarial training increase the model's robustness to MI attacks, effectively balancing privacy, efficiency and utility.

\textbf{Dual Gradient Pruning (DGP).} Xue et al.~\citep{Xue_Hu_Zhao_Zhang_Hu_Sun_Yao_2024} investigated privacy challenges posed by GIAs in collaborative learning. Given that existing defense methods (e.g., differential privacy and cryptography) struggle to achieve a balance between privacy, utility and efficiency, the authors proposed DGP, which enhances the traditional gradient pruning method by improving the efficiency of communication in collaborative learning systems while providing stronger privacy guarantees. The method aims to reduce the risk of data recovery by selectively pruning the gradient to resist powerful GIAs without affecting model performance. Through both theoretical analysis and extensive experiments, the authors demonstrated that DGP not only provides strong GIA defense, but also reduces the communication cost significantly, making it an efficient solution for privacy-preserving collaborative learning.

\textbf{Guardian.} Fan et al.~\citep{Fan_Liu_Chen_Wang_Qiu_Zhou_2024} developed Guardian to combat gradient leakage by optimising two theoretically derived metrics: the performance maintenance metric and privacy protection metric. These metrics are used to generate transformed gradients that minimize privacy leakage while maintaining model accuracy. The authors introduced an innovative initialization strategy to speed up the production of transformed gradients, thus increasing the utility of Guardian. The authors provided theoretical convergence guarantees and conducted extensive experiments on a variety of tasks, including tasks using visual transformer architectures, showing that Guardian can effectively withstand state-of-the-art attacks with no significant loss of accuracy. Guardian's defense ability even in strict white-box scenarios with Bayesian optimal adversaries highlights its potential for real-world applications.

\textbf{Pruned Frequency-based Gradient Defense (pFGD).} Palihawadana et al.~\citep{Palihawadana_Wiratunga_Kalutarage_Wijekoon_2023} presented pFGD,  which incorporates frequency transformation techniques such as discrete cosine transform and gradient pruning methods to strengthen privacy protection. The experimental results on the MNIST dataset show that pFGD substantially reduces the risk from gradient inversion attacks while exhibiting resilience and robustness.

\textbf{Deep Gradient Compression (DGC).} Lin et al.~\citep{Lin_Han_Mao_Wang_Dally_2018} proposed DGC to keep model accuracy while minimizing communication bandwidth. This approach reduces gradient exchange by 99.9\% via strategies such as momentum correction and local gradient clipping, and proves its effectiveness on various datasets, including ImageNet and Penn Treebank, with compression rates of up to 600 times. Although compression strategies like DGC contribute to scalability and efficiency, they also raise questions about their potency to affect the vulnerability of adversaries, as compression may change the attack surface of MI strategies. Thus, the intersection of gradient compression and MI remains an area of interest in protecting data privacy.

\subsection{Gradient Perturbation/Obfuscation}

Gradient perturbation and obfuscation can protect sensitive data during model training. Perturbation introduces small variations or noise into the gradient, masking sensitive data while retaining utility, while obfuscation obscures the gradient, making it difficult to interpret. These techniques are vital to defending against MI attacks. Some gradient perturbation and obfuscation methods are reviewed below.

\textbf{Model Fragmentation, Shuffle and Aggregation.} Masuda et al.~\citep{Masuda_Kita_Koizumi_Takemasa_Hasegawa_2021} used model fragmentation, shuffling, and aggregation to tackle MI attacks without compromising model accuracy. In this work, each participant fragments, shuffles and aggregates models before sharing them, thus making it difficult for adversaries to reconstruct data. The method maintains the integrity of the shared model in the defense of MI attacks, which is an important step forward in enhancing the privacy of FL and reflects the ongoing requirement for robust and effective privacy-preserving strategies in joint learning environments.

\textbf{Autoencoder-based Compression.} Chen et al.~\citep{Chen_Abrahamyan_Sahli_Deligiannis_2024} presented an adaptive autoencoder-based approach to compressing and perturbing model parameters before sharing them with a server, thus improving privacy without sacrificing model performance. In contrast to traditional defense methods that introduce noise or sparse gradients (which might prevent model convergence and increase communication costs), this approach allows clients to gain a compressed representation of their local model parameters in just a few iterations. The empirical evaluation shows that the proposed method reduces the communication rate by 4.1 times compared to the joint averaging approach, while retaining almost the same model performance as unmodified local updates. Furthermore, the results show that the method is effective in preventing information leakage, demonstrating the potential of autoencoder-based compression techniques for achieving beneficial trade-offs between privacy preservation, model efficiency, and communication costs in joint learning environments.

\textbf{GradPrivacy.} Lu et al.~\citep{Lu_Xue_Wan_Li_Zhang_Hu_2023} worked on a critical gap in privacy protection in collaborative learning systems, especially defenses against GIAs. While prior research mainly focused on untrained models, the authors emphasized that trained models carry more information and are equally vulnerable to attacks, and thus must be prioritized for protection. To reduce privacy risks, the authors proposed GradPrivacy to protect privacy at all stages of collaborative learning while not sacrificing model performance. GradPrivacy consists of two main components: an amplitude perturbation module and a deviation correction module. The former is used to perturb the gradient parameters associated with sensitive features in order to impede gradient inversion. The latter aims to correct deviations in model updates so that the model maintains its accuracy. Through extensive evaluation, the authors demonstrated that GradPrivacy achieves an excellent balance between privacy and accuracy, outperforming existing defenses by providing robust protection for well-trained models in collaborative learning environments.

\textbf{Quantization Enabled FL.} Ovi et al.~\citep{Ovi_Dey_Roy_Gangopadhyay_2023} showed that traditional defense mechanisms, such as differential privacy, homomorphic encryption, and gradient pruning, tend to suffer from drawbacks such as complex key generation, degraded performance, and difficulty in selecting the optimum pruning ratio. To address these drawbacks, the authors designed a joint learning scheme with hybrid quantization, in which different layers of a deep learning model are quantized with different accuracies and modes. Their experimental evaluation shows that this approach greatly improves robustness against iteration- and recursion-based gradient inversion attacks, while keeping strong performance on multiple benchmark datasets.

\subsection{Differential Privacy}

Differential privacy (DP), which uses the idea of noise to limit information leakage, has emerged as a foundational framework for mitigating privacy risks posed by MI attacks. By introducing carefully calibrated noise into the training process or model output, DP limits the possibility of reconstructing sensitive information by ensuring that the presence or absence of any individual data point in the training set has a minimum impact on model predictions. In the setting of MI attacks, DP-based defense mechanisms focus on techniques such as gradient noise addition, output perturbation, and privacy budget allocation to obscure the link between training data and model response. Despite theoretical guarantees, these defenses often confront challenges in balancing privacy with model utility, especially in areas that require high accuracy or deal with complex data types. This section explores some DP-related approaches to defending against MI attacks.

\textbf{Augmented DP.} While transparency in deep learning enhances trust, accountability, and fairness, traditional methods such as DP often fail to fully protect against MI attacks, particularly with high accuracy. Alufaisan et al.~\citep{Alufaisan_Kantarcioglu_Zhou_2020} introduced an improved DP method to create transparent, accurate models that are resistant to MI attacks, thus bridging this gap. By improving the balance between privacy and utility, this approach provides strong privacy guarantees without severely compromising model accuracy, thus addressing a common trade-off in privacy-preserving deep learning. This research highlights the significance of designing methods that are both privacy-preserving and transparent.

\textbf{Class-level and Subclass-level DP.} Zhang et al.~\citep{Zhang_Ma_Xiao_Lou_Xiong_2020} indicated that standard record-level DP, while popular for privacy-preserving deep learning, lacks robustness against MI attacks. The authors enhanced existing MI attack techniques by demonstrating their ability to reconstruct training images from deep learning models, stressing the need for more fine-grained privacy protection mechanisms. To this end, the authors proposed a novel class-level and subclass-level DP scheme designed to provide quantifiable privacy guarantees, specifically aimed at defeating MI attacks. The experiments show that the proposed scheme successfully strikes a balance between privacy protection and model accuracy, driving the field toward more effective defenses against MI attacks in deep learning.

\textbf{DP in Healthcare Models.} Krall et al.~\citep{Krall_Finke_Yang_2020} developed a gradient-based scheme to preserve DP in healthcare modeling, especially in an intensive care unit setting. This scheme aims to mitigate the adversary's ability to infer sensitive patient attributes by applying DP techniques during gradient descent to reduce the risk of MI attacks. The experimental results show that the scheme reduces the risk of MI while maintaining high model accuracy, highlighting its potential in the healthcare domain where data protection is critical.

\textbf{Local DP.} Li et al.~\citep{Li_Yu_Cheng_Yan_Zhang_2024} studied the important issue of gradient privacy in joint learning, where an attacker can infer local data from uploaded gradients. To address this challenge, the authors proposed a privacy-enhancing approach that integrates local DP, parameter sparsification, and weighted aggregation, which is particularly suitable for cross-silo settings. Their approach uses DP by adding noise to local parameters before uploading, which achieves local DP while adjusting the privacy budget dynamically to balance noise and model accuracy. In addition, the authors introduced the Top-K method to optimise communication costs based on the varying capabilities of clients and used weighted aggregation to augment the robustness of the privacy framework. The experimental results show that this approach effectively balances privacy, accuracy, communication cost, and robustness.

\subsection{Cryptographic Encryption}
Recent advances in cryptography have heightened the security of federated and distributed deep learning systems against MI attacks. By incorporating methods such as secure gradient aggregation, perceptual hashing, and homomorphic encryption, researchers are tackling critical privacy challenges while maintaining model accuracy and efficiency. This section analyzes these innovative approaches to demonstrate their potential to strengthen data protection in collaborative learning environments.

\textbf{Secure Aggregation.} Yang et al.~\citep{Yang_Yang_Huang_Martínez_López_Chen_2023} devised a secure aggregation approach to thwarting MI attacks in FL environments. The method consists of encrypting the gradient before sharing it, preventing the adversary from using the gradient information to reconstruct private data. To improve efficiency, the authors developed a new method for producing shared keys, where each client establishes keys with a subset of other clients instead of all clients in the system. Simulation results show that the proposed method is effective against attacks initiated by honest but curious parameter servers. 

\textbf{Perceptual Hashing.} Prakash et al.~\citep{Prakash_Ding_Li_Errapotu_Pei_Pan_2020} proposed a privacy-preserving approach to mitigate MI attacks using perceptual hashing of training images. Their approach converts parts of each training image into a hashed form and then uses these perceptually hashed images to train facial recognition models, thereby effectively mitigating the risk of reconstructing the original image in an inversion attack. This method retains a high level of classification accuracy while providing a boost in privacy, as the hashed image, instead of the original image, is returned in the adversarial scenario.

\textbf{Certificateless Additively Homomorphic Encryption (CAHE).} Based on learning with errors, Antwi-Boasiako et al.~\citep{Boasiako_Zhou_Liao_Dong_2023} designed CAHE to resist MI attacks. Their work eliminates the need for trusted third party and reduces reliance on centralized authorities, thus providing better privacy protection, while addressing the high communication cost associated with transmitting cryptographic gradients in distributed deep learning systems. Their results show that the accuracy of the CAHE-based model remains high (97.20\%) while ensuring privacy protection of participant data. Their solution also employs a partial gradient sharing algorithm, which improves communication efficiency and achieves an accuracy of 97.17\% and 97.12\% in the case of upmost gradient selection and random gradient selection, respectively. 

\subsection{Model/Architecture Enhancement} \label{enhance}

Recent advances in privacy-preserving machine learning have highlighted the importance of improving model architectures to tackle security vulnerabilities, especially in federated and collaborative learning environments. This section goes over model/architecture-based approaches to defending against MI attacks.

\textbf{PRECODE.} Scheliga et al.~\citep{Scheliga_Mäder_Seeland_2022} introduced PRECODE, a privacy-enhancing module intended to protect gradient leakage in collaborative neural network training. In contrast to traditional gradient perturbation techniques that tend to degrade model performance or increase training complexity, PRECODE involves random sampling using variational modeling to effectively protect client data from gradient inversion attacks. The module is a flexible extension for a variety of model architectures, providing robust protection without compromising training efficiency or accuracy. Through extensive testing on multiple model architectures and datasets, the authors showed that PRECODE reduces the attack success rate to 0\%, outperforming existing defense mechanisms, and demonstrating that PRECODE is a promising and effective solution to enhancing privacy protection in distributed learning environments without the trade-offs typically associated with gradient perturbation techniques.

\textbf{Secure Convolutional Neural Networks (SecCNN).} Liu et al.~\citep{Liu_Shen_Chen_Chen_2024} proposed SecCNN to integrate an upsampling layer into the CNN, offering an intrinsic defense against gradient inversion attacks. This approach also leverages rank analysis to increase security without compromising model accuracy or adding computational cost. The work advances current defense strategies for FL, highlighting the potential of model architecture modifications as an effective tool to combat MI attacks while maintaining computational efficiency.

\textbf{Variational Encoder-based Personalized FL (RVE-PFL).} Issa et al.~\citep{Issa_Moustafa_Turnbull_Choo_2024} devised the RVE-PFL approach to alleviating  MI attacks while preserving model utility. The personalized variant encoder architecture assures privacy of heterogeneous data across clients and efficiently aggregates data at the server level, differentiating between adversarial settings and legitimate operations. Research shows that privacy-preserving techniques in FL typically weaken model performance, whereas RVE-PFL offers significant improvements in both privacy and utility.

\textbf{ResSFL.} Li et al.~\citep{Li_Rakin_Chen_He_Fan_Chakrabarti_2022} designed a two-step framework called ResSFL to resolve the vulnerability of split FL against MI attacks. The ResSFL framework utilizes attacker-aware training in combination with a bottleneck layer to exploit an MI-resistant feature extractor, which is subsequently used by the client for secure and efficient collaborative learning. This approach guarantees robustness against MI attacks during the critical training phase and significantly improves resilience with minimal computational overhead. An extensive evaluation on datasets such as CIFAR-100 validates the efficacy of ResSFL, which achieves an excellent trade-off between accuracy and resistance to information interference when compared to baseline and contemporary approaches.

\subsection{Miscellaneous MI Defenses}

\textbf{Transfer Learning-based Defense against Model Inversion (TL-DMI).} Many existing defense methods rely on regularization techniques, which often reduce model performance by conflicting with training objectives. Ho et al.~\citep{Ho_Hao_Chandrasegaran_Nguyen_Cheung_2024} proposed a simple yet effective defense method, called TL-DMI, to counter MI attacks using transfer learning~(TL). This approach reduces the attacker's ability to reconstruct private data by limiting the number of layers that encode sensitive information in the training dataset. Through analyzing Fisher information, the authors theoretically justified this approach, demonstrating that TL can effectively reduce MI attack success. Extensive experiments prove that TL-DMI is invulnerable to MI attacks and achieves robust privacy protection without impairing the utility of the model. The simplicity and effectiveness of TL-DMI make it a compelling defense strategy for deep learning privacy protection.

\textbf{Bilateral Dependency Optimisation (BiDO).} Peng et al.~\citep{Peng_Liu_Zhang_Lan_Ye_Liu_Han_2022} designed BiDO to defend MI attacks recovering the training data from the classifier and leading to privacy breaches. Traditional defense strategies focus on minimizing the dependency between inputs and outputs during training, however, this conflicts with the need to maximize this dependency for accurate classification, thus necessitating a trade-off between the strength of the defense and the utility of the model. BiDO addresses this problem by minimizing the dependency between potential representations and inputs, and simultaneously maximizing the dependency between potential representations and outputs. The authors proposed two implementations of BiDO: BiDO-COCO (using constrained covariance) and BiDO-HSIC (based on the Hilbert-Schmidt independence criterion). The experiments show that BiDO tackles MI attacks with minimal impact on classification accuracy, offering a new and effective approach to balancing privacy and model performance.

\textbf{Mutual Information Regularization-based Defense (MID).} Accredited to Wang et al.~\citep{Wang_Zhang_Jia_2021}, MID controls the captured information about the model input in the prediction and reduces the adversary's ability to reconstruct privacy attributes. The authors showed that MID achieves superior performance on a range of models and datasets, providing a theoretically sound alternative to traditional methods such as differential privacy.

\subsection{Summary}

Defense strategies against MI attacks have evolved into diverse approaches to dealing with privacy vulnerabilities while attempting a good balance between efficiency, practicality, and robustness. These approaches are organized into six categories (see Sections~\ref{perturbation}-\ref{enhance}). Some insights are: feature perturbation techniques (e.g., Crafter and Soteria) distort sensitive data while maintaining accuracy; gradient perturbation strategies (e.g., GradPrivacy) show promise in obfuscating data in collaborative learning processes; DP remains crucial, and innovations such as class-level DP extend privacy guarantees; encryption techniques, such as certificateless homomorphic encryption, secure aggregation, and perceptual hashing, provide robust data protection by securing shared information; and model-enhancing techniques such as PRECODE and ResSFL protect privacy through architectural enhancements that defend against attacks without sacrificing performance.

Despite recent progress, it is challenging to achieve a balanced trade-off between privacy, utility, and computational costs, especially to counter advanced threats~\citep{Scheliga_Mäder_Seeland_2023} and address security concerns in IoT and edge computing environments. Most of the MI defense mechanisms discussed earlier, except for gradient pruning methods such as PATROL~\citep{Ding_Zhang_Pan_Yuan_2024}, DGP~\citep{Xue_Hu_Zhao_Zhang_Hu_Sun_Yao_2024}, Guardian~\citep{Fan_Liu_Chen_Wang_Qiu_Zhou_2024}, and DGC~\citep{Lin_Han_Mao_Wang_Dally_2018}, fail to consider the trade-off between efficiency and privacy preservation in IoT and edge computing settings. These environments are often constrained by limited computational resources, making traditional, resource-intensive security defenses impractical. Gradient pruning methods selectively prune gradients to enhance privacy and communication efficiency while maintaining model performance. For instance, DGC~\citep{Lin_Han_Mao_Wang_Dally_2018} reduces gradient exchange by 99.9\% and still preserves model accuracy. However, nearly all existing gradient pruning methods have yet to be tested on real IoT or edge computing devices. Therefore, there is a pressing need for practical pruning strategies and lightweight MI defense mechanisms (e.g., efficient encryption techniques) to mitigate vulnerabilities while ensuring a feasible trade-off between defense performance and cost effectiveness (e.g., computational efficiency)  in IoT and edge computing environments.

\section{Evaluation Metrics in MI Attacks and Defenses}
\label{sec:Evaluation Metrics}

In this section, the most common evaluation metrics used in MI attacks and defenses are discussed, along with detailed explanations for each metric~\citep{Issa_Moustafa_Turnbull_Choo_2024, Shi_Kotevska_Reshniak_Singh_Raskar_2024, Zhang_Cheng_Shen_Ribeiro_An_Chen_Zhang_Li_2025}. The provided evaluation metrics range from simple pixel-based comparisons to complex feature and latent-space-based evaluations.

\

\noindent\textbf{Mean Squared Error (MSE)}~\citep{Bishop_Nasrabadi_2006, Dodge_2008}: MSE is a standard metric for evaluating the quality of reconstructions in MI attacks. It quantifies the average squared difference between the reconstructed data and the actual target data. A lower MSE indicates that the reconstructed data is closer to the original data, signifying a more successful attack. MSE is particularly useful in image reconstruction or attribute inference tasks, where fidelity to the original data is critical. MSE can be computed as:

\begin{equation}
\text{MSE} = \frac{1}{N} \sum_{i=1}^N \left( x_i - y_i \right)^2
\end{equation}
where \(x_i\) represents the predicted value and \(y_i\) the ground truth value. Smaller MSE values indicate better performance.

\

\noindent \textbf{Learned Perceptual Image Patch Similarity (LPIPS)}~\citep{Zhang_Isola_Efros_Shechtman_Wang_2018}: LPIPS measures perceptual similarity by comparing representations in neural network feature spaces. It captures semantic and perceptual nuances better than traditional metrics. LPIPS is particularly effective for evaluating image reconstructions in visually complex tasks. LPIPS is expressed as:
\begin{equation}
\text{LPIPS}(\mathbf{x}, \mathbf{y}) = \sum_{l} \frac{1}{H_l W_l} \sum_{h,w} \| \phi_l^h(\mathbf{x}) - \phi_l^h(\mathbf{y}) \|^2
\end{equation}
where \( \phi_l(\cdot) \) represents features extracted from a pre-trained neural network at layer~\(l\); and \(H_l\) and \(W_l\) are spatial dimensions of features at layer~\(l\).

\ 

\noindent\textbf{Peak Signal-to-Noise Ratio (PSNR)}~\citep{Lin_Dong_Xue_2005, Wang_Bovik_Sheikh_Simoncelli_2004}: PSNR is a traditional measure for quantifying the quality of reconstruction in lossy compression. It compares the maximum possible signal power to the power of noise affecting the fidelity. PSNR is obtained by:
\begin{equation}
\text{PSNR} = 10 \cdot \log_{10} \left( \frac{\text{MAX}^2}{\text{MSE}} \right)
\end{equation}
where \(\text{MAX}\) denotes the maximum possible pixel value (e.g., 255 for 8-bit images), and \(\text{MSE}\) the Mean Squared Error between the original and reconstructed images

\ 

\noindent\textbf{Structural Similarity Index Measure (SSIM)}~\citep{Wang_Simoncelli_Bovik_2003, Wang_Bovik_Sheikh_Simoncelli_2004}: Unlike traditional error measures like MSE, SSIM is a perceptual metric for assessing image quality based on structural similarity. It measures the similarity between two images by comparing their luminance, contrast, and structural information. SSIM assumes that human visual systems are adapted to extract structural information from visual scenes. SSIM is calculated by:
\begin{equation}
\text{SSIM}(x, y) = \frac{(2\mu_x \mu_y + C_1)(2\sigma_{xy} + C_2)}{(\mu_x^2 + \mu_y^2 + C_1)(\sigma_x^2 + \sigma_y^2 + C_2)}
\end{equation}
where \( \mu_x, \mu_y \) are mean intensities of images \(x\) and \(y\), respectively; \( \sigma_x^2, \sigma_y^2 \) are variances of images~\(x\) and \(y\), respectively; \( \sigma_{xy} \) is the covariance between images~\(x\) and \(y\); and \( C_1, C_2 \) are stabilizing constants.

\ 

\noindent\textbf{fréchet inception distance (FID)}~\citep{Heusel_Ramsauer_Unterthiner_Nessler_Hochreiter_2017}: FID measures the similarity between feature distributions of real and generated images using the Wasserstein-2 distance. FID is a widely used metric to evaluate the quality of generated images in tasks such as GANs. FID is written as:

\begin{equation}
\text{FID} = \|\mu_x - \mu_y\|_2^2 + \text{Tr}(\Sigma_x + \Sigma_y - 2 (\Sigma_x \Sigma_y)^{1/2})
\end{equation}
where \( \mu_x, \mu_y \) are the mean vectors of features extracted from real and generated images, respectively; \( \Sigma_x, \Sigma_y \) are covariance matrices of features extracted from real and generated images, respectively; and \( \text{Tr}(\Sigma_x + \Sigma_y - 2 (\Sigma_x \Sigma_y)^{1/2}) \) computes the trace of the resulting matrix from the covariance terms, capturing the spread and relationships between the real and generated data distributions.

\ 

\noindent \textbf{Feature Similarity Index (FSIM)}~\citep{Zhang_Zhang_Mou_Zhang_2011, Shi_Kotevska_Reshniak_Singh_Raskar_2024}: FSIM is a recent metric that uses phase congruency (a measure of feature perception) and gradient magnitude to evaluate image quality. FSIM captures human perception-based similarity and can be computed as:

\begin{equation}
\text{FSIM} = \frac{\sum_{p \in \Omega} \text{PC}_m(p) \cdot \text{S}(p)}{\sum_{p \in \Omega} \text{PC}_m(p)}
\end{equation}
where \( \Omega \) is a set of pixels in the image; \( \text{PC}_m(p) \) is the phase congruency value at pixel~\(p\) (perceptually motivated feature); and \( \text{S}(p) \) is the similarity of gradient magnitude and phase congruency between two images.

\ 

\noindent \textbf{Absolute Variation Distance (AVD)}~\citep{Papadopoulos_Satsangi_Eloul_Pistoia_2024}: AVD is a metric for evaluating data recovery and information leakage, specifically in the context of FL and inversion attacks. It provides a method to compare the similarity between a reconstructed (or attacked) image and its original counterpart by analyzing their spatial gradients. AVD is expressed as:
\begin{equation}
\text{AVD}(\mathbf{v}_{\text{source}}, \mathbf{v}_{\text{target}}) = \| |\nabla \mathbf{v}_{\text{source}}||\nabla \mathbf{v}_{\text{target}}| \| + \| |\nabla^2 \mathbf{v}_{\text{source}}| - |\nabla^2 \mathbf{v}_{\text{target}}| \|
\end{equation}
where \(\nabla \mathbf{v} = \frac{\partial \mathbf{v}}{\partial i} + \frac{\partial \mathbf{v}}{\partial j}\) is the spatial gradient; \(\nabla^2 \mathbf{v} = \frac{\partial^2 \mathbf{v}}{\partial i^2} + \frac{\partial^2 \mathbf{v}}{\partial j^2}\) is the second-order spatial gradient; and \(\mathbf{v}_{\text{source}}\) and \(\mathbf{v}_{\text{target}}\) are source and target images, treated as 2D arrays of pixel values.

\ 

\noindent \textbf{Relative Data Leakage Value (RDLV)}~\citep{Hatamizadeh_Yin_Molchanov_Myronenko_Li_Dogra_Feng_Flores_Kautz_Xu_2023}: RDLV is a metric for quantifying and comparing the degree of data leakage in gradient inversion attacks within FL systems. It is particularly useful for evaluating privacy risks across multiple clients under varying privacy-preserving configurations. RDLV is defined as:
\begin{equation}
\text{RDLV} =
\frac{\text{SSIM}(T_i, I_i) - \text{SSIM}(T_i, P)}{\text{SSIM}(T_i, P)}
\end{equation}
where \(T_i\) is the original training image; \(I_i\) the reconstructed image obtained from the gradient inversion attack; and \(P\) is the prior used in the gradient inversion attack (e.g., an initialization image for reconstruction). 

\ 

\noindent \textbf{Image Identifiability Precision (IIP)}~\citep{Yin_Mallya_Vahdat_Alvarez_Kautz_Molchanov_2021}: IIP quantifies the extent of ``image-specific" information disclosed through gradient inversion. It assesses how effectively a specific image can be identified based solely on its reconstructed version among other similar images in the original dataset. Numerically, IIP is determined as the proportion of exact matches between an original image and its closest neighbor in the reconstructed set.


\section{Datasets for MI Attack Research}
\label{sec:Datasets}

Existing studies on MI attacks and defenses employ diverse datasets to evaluate attack efficacy and defense mechanisms. These datasets contain different data types and complexities, facilitating a comprehensive assessment of privacy vulnerabilities and protections in machine learning models. Below is a list of commonly used datasets in MI-related research.

\ 

\noindent \textbf{MNIST -- Modified National Institute of Standards and Technology Database}~\citep{LeCun_1998}: MNIST is a large database of handwritten digits (0 through 9), consisting of 60,000 training images and 10,000 test images, each represented as 28x28 grayscale pixels. Works using this database include~\citep{Wang_Fu_Li_Khisti_Zemel_Makhzani_2021, Zhao_Mopuri_Bilen_2020, Sun_Li_Wang_Yang_Li_Chen_2021, Chen_Kahla_Jia_Qi_2021, Zhao_Zhang_Xiao_Lim_2021, Titcombe_Hall_Papadopoulos_Romanini_2021, Peng_Liu_Zhang_Lan_Ye_Liu_Han_2022, Scheliga_Mäder_Seeland_2023, Nguyen_Chandrasegaran_Abdollahzadeh_Cheung_2023, Ren_Deng_Xie_Ma_Ma_2023, Xu_Zhang_Ding_Wang_2024, Issa_Moustafa_Turnbull_Choo_2024, Liu_Wang_Ren_Wang_Guo_Qin_Liu_2024, Li_Yu_Cheng_Yan_Zhang_2024, Wang_Guo_Deng_Zhang_Fang_2024, Gao_Zhu_Ye_Zhou_2024, Fan_Liu_Chen_Wang_Qiu_Zhou_2024, Wu_Hayat_Zhou_Harandi_2024, Wang_Guo_Xie_Qi_2022, Niu_Wang_Zhang_Guo_Cao_Li_2023, Liu_Li_Gao_Xie_Zhao_2023, Wang_Lee_Lei_2023, Wan_Du_Yuan_Yang_Chen_Xu_2023, Yang_Hang_Ding_Li_Liang_Liu_2024, Xiao_Li_Li_2024, Qi_Wang_Huang_2024, Zhu_Huang_Xie_2024}.

\ 

\noindent \textbf{F-MNIST -- Fashion MNIST}~\citep{Xiao_Rasul_Vollgraf_2017}: This dataset comprises 28x28 grayscale images of 70,000 fashion products from 10 categories, with 7,000 images per category. The training set has 60,000 images and the test set has 10,000 images. Works using this database include~\citep{Gao_Zhang_Guo_Zhang_Xiang_Qiu_Wen_Liu_2023, Chen_Abrahamyan_Sahli_Deligiannis_2024, Issa_Moustafa_Turnbull_Choo_2024, Wang_Guo_Xie_Qi_2022, Xiao_Li_Li_2024, Qi_Wang_Huang_2024}.

\ 

\noindent \textbf{CIFAR-10 -- Canadian Institute for Advanced Research - 10 classes} \citep{Krizhevsky_Nair_Hinton_2010}: This database consists of 60,000 color images~(32x32) in 10 classes, with 6,000 images per class, divided into 50,000 training and 10,000 test images. Works using this database include~\citep{Sun_Li_Wang_Yang_Li_Chen_2021, Chen_Kahla_Jia_Qi_2021, Peng_Liu_Zhang_Lan_Ye_Liu_Han_2022, Scheliga_Mäder_Seeland_2022, Lu_Xue_Wan_Li_Zhang_Hu_2023, Scheliga_Mäder_Seeland_2023, Nguyen_Chandrasegaran_Abdollahzadeh_Cheung_2023, Ren_Deng_Xie_Ma_Ma_2023, Xu_Zhang_Ding_Wang_2024, Hu_Wang_Dong_Xue_2024, Chen_Abrahamyan_Sahli_Deligiannis_2024, Issa_Moustafa_Turnbull_Choo_2024, Zhou_Zhu_Ye_Zhou_Zhao_2024, Li_Yu_Cheng_Yan_Zhang_2024, Wang_Guo_Deng_Zhang_Fang_2024, Gao_Zhu_Ye_Zhou_2024, Fan_Liu_Chen_Wang_Qiu_Zhou_2024, Wu_Hayat_Zhou_Harandi_2024, Xue_Hu_Zhao_Zhang_Hu_Sun_Yao_2024, Wang_Guo_Xie_Qi_2022, Jiang_Wang_Valls_Ko_Lee_Leung_Tassiulas_2022, Yang_Feng_Fang_Shao_Tang_Xia_Lu_2022, Wan_Xu_Liu_Ren_Fan_Tang_2022, Zhang_Tianqing_Ren_Xiong_Choo_2023, Niu_Wang_Zhang_Guo_Cao_Li_2023, Zhu_Shi_Luo_Wang_Peng_Fan_Letaief_2023, Liu_Li_Gao_Xie_Zhao_2023, Wan_Du_Yuan_Yang_Chen_Xu_2023, Chu_Yang_Laoutaris_Markopoulou_2023, Fan_Chen_Wang_Li_Zhou_Huang_2023, Noorbakhsh_Zhang_Hong_Wang_2024, Yang_Hang_Ding_Li_Liang_Liu_2024, Qi_Wang_Huang_2024, Zhu_Huang_Xie_2024}.

\ 

\noindent \textbf{CIFAR-100}~\citep{Krizhevsky_Hinton_2009}: This dataset is similar to CIFAR-10 but includes 100 distinct classes, each comprising 600 images. For each class, there are 500 images for training and 100 images for testing. The 100 classes are organized into 20 superclasses, with each image labeled at two levels: a ``fine" label indicating its specific class and a ``coarse" label representing its broader superclass. Works using this database include~\citep{Zhao_Mopuri_Bilen_2020, Gao_Guo_Zhang_Qiu_Wen_Liu_2021, Scheliga_Mäder_Seeland_2022, Lu_Xue_Wan_Li_Zhang_Hu_2023, Ren_Deng_Xie_Ma_Ma_2023, Hu_Wang_Dong_Xue_2024, Chen_Abrahamyan_Sahli_Deligiannis_2024, Zhou_Zhu_Ye_Zhou_Zhao_2024, Xue_Hu_Zhao_Zhang_Hu_Sun_Yao_2024, Wang_Guo_Xie_Qi_2022, Zhang_Tianqing_Ren_Xiong_Choo_2023, Wan_Du_Yuan_Yang_Chen_Xu_2023, Fan_Chen_Wang_Li_Zhou_Huang_2023, Noorbakhsh_Zhang_Hong_Wang_2024, Wang_Hugh_Li_2024, Zhu_Huang_Xie_2024}.

\ 

\noindent \textbf{LFW -- Labeled Faces in the Wild}~\citep{Huang_Mattar_Berg_Learned_Miller_2008}: This dataset is a collection of face photographs aimed at advancing research in unconstrained face recognition. Developed and maintained by researchers at the University of Massachusetts, Amherst, the dataset comprises 13,233 images of 5,749 individuals, with faces detected and centered using the Viola-Jones face detector. These images were sourced from the web with 1,680 individuals having two or more unique photographs included in the dataset. Works using this database include~\citep{Zhao_Mopuri_Bilen_2020, Zhang_Tianqing_Ren_Xiong_Choo_2023, Cui_Meerza_Li_Liu_Zhang_Liu_2023}.

\ 

\noindent \textbf{CelebA -- CelebFaces Attributes Dataset}~\citep{Liu_Luo_Wang_Tang_2015}: This dataset is a large-scale collection of over 200,000 celebrity images, each annotated with 40 binary attributes. It includes 10,177 unique identities and 202,599 face images, with annotations for five facial landmarks and 40 attributes per image. The dataset is immensely diverse, with variations in pose and background clutter, making it suitable for a range of computer vision tasks such as facial attribute recognition, face recognition, face detection, landmark localization, and face editing or synthesis. Works using this database include~\citep{Wang_Fu_Li_Khisti_Zemel_Makhzani_2021, Chen_Kahla_Jia_Qi_2021, Zhao_Zhang_Xiao_Lim_2021, Peng_Liu_Zhang_Lan_Ye_Liu_Han_2022, Gao_Zhang_Guo_Zhang_Xiang_Qiu_Wen_Liu_2023, Nguyen_Chandrasegaran_Abdollahzadeh_Cheung_2023, MaungMaung_Kiya_2023, Nguyen_Chandrasegaran_Abdollahzadeh_Cheung_2024, Dibbo_Breuer_Moore_Teti_2024, Liu_Wang_Ren_Wang_Guo_Qin_Liu_2024, Wu_Hayat_Zhou_Harandi_2024, Jiang_Wang_Valls_Ko_Lee_Leung_Tassiulas_2022, Li_Zhang_Liu_Liu_2022, Cui_Meerza_Li_Liu_Zhang_Liu_2023, Yu_Qiu_Fang_Chen_Yu_Wang_Xia_Xu_2024}.

\ 

\noindent \textbf{ImageNet}~\citep{Deng_Dong_Socher_Li_Li_Fei-Fei_2009}: ImageNet is a vast visual database created to support research in visual object recognition. It includes over 14 million hand-annotated images, with objects labeled to specify their contents, and bounding boxes provided for at least one million images. Spanning more than 20,000 categories, such as ``balloon" or ``strawberry," each category typically contains several hundred images. ImageNet is extensively used for training and benchmarking deep learning models. Works using this database include~\citep{Gao_Zhang_Guo_Zhang_Xiang_Qiu_Wen_Liu_2023, Fang_Chen_Wang_Wang_Xia_2023, MaungMaung_Kiya_2023, Li_Zhang_Liu_Liu_2022, Zhu_Huang_Xie_2024}.

\ 

\noindent \textbf{FFHQ -- Flickr-Faces-HQ}~\citep{Karras_2019}: This is a high-resolution image dataset featuring 70,000 PNG images of human faces at a resolution of 1024×1024 pixels. The dataset exhibits significant diversity in age, ethnicity, and backgrounds, along with extensive coverage of accessories such as eyeglasses, sunglasses, and hats, making it well-suited for applications in face-related research. Works using this database include~\citep{Nguyen_Chandrasegaran_Abdollahzadeh_Cheung_2023, Fang_Chen_Wang_Wang_Xia_2023, Nguyen_Chandrasegaran_Abdollahzadeh_Cheung_2024}.

\ 

\noindent \textbf{ChestX-ray8}~\citep{Wang_Peng_Lu_Lu_Bagheri_Summers_2017}: This is a medical imaging dataset containing 108,948 frontal-view X-ray images from 32,717 patients, collected between 1992 and 2015. It includes eight common disease labels extracted from radiological reports using NLP techniques, making it a valuable resource for medical image analysis and disease diagnosis research. Works using this database include~\citep{Wang_Fu_Li_Khisti_Zemel_Makhzani_2021, Chen_Kahla_Jia_Qi_2021}.

\ 

\noindent \textbf{UBMD -- UCI Bank Marketing Dataset}~\citep{Moro_Cortez_Rita_2014}: This dataset is designed to predict the likelihood of clients subscribing to deposits. It consists of 41,188 instances with 17-dimensional bank data, providing a comprehensive basis for analysis and modeling. Works using this database include~\citep{Yang_Feng_Fang_Shao_Tang_Xia_Lu_2022}.

\ 

\noindent \textbf{LDC -- Lesion Disease Classification}~\citep{Tschandl_Rosendahl_Kittler_2018}: LDC has 8,000 training images and 2,000 test images of skin lesions, intended for the classification of various skin diseases. Works using this database include~\citep{Yang_Feng_Fang_Shao_Tang_Xia_Lu_2022}.

\ 

\noindent \textbf{AT\&T -- The Database of Faces}: This dataset contains 400 grayscale images of 40 subjects, each with 10 images taken under varying conditions such as lighting, facial expressions, and details (e.g., glasses). Each image is of 92x112 pixels with 256 gray levels, organized in 40 directories by subject, totaling 4.5 MB in size. Originally created between 1992 and 1994 for a face recognition project, this dataset provides a consistent dark background and frontal face orientation, making it ideal for facial recognition research. Works using this database include~\citep{Melis_Song_De_Cristofaro_Shmatikov_2019}. 

\ 

\noindent \textbf{SVHN -- Street View House Numbers}~\citep{Netzer_Wang_Coates_Bissacco_Wu_Ng_2011}: SVHN is a benchmark dataset for digit classification, consisting of 600,000 32×32 RGB images of printed digits (0–9) cropped from house number plates. The images center around the target digit while retaining nearby digits and other visual elements. This dataset is divided into three subsets: training, testing, and an additional set with 530,000 less challenging images to assist in the training process. Works using this database include~\citep{Fan_Chen_Wang_Li_Zhou_Huang_2023}.




\section{Challenges and Future Research Directions}
\label{sec:Open Challenges}

While substantial progress has been made in understanding and mitigating MI attacks, many challenges remain. This section discusses some of these challenges and corresponding future research directions.

\subsection{Balancing Privacy and Model Utility}
Balancing privacy and model utility is a pressing challenge in developing defenses against MI attacks. Existing methods (e.g.,~\citep{Ho_Hao_Chandrasegaran_Nguyen_Cheung_2024, Peng_Liu_Zhang_Lan_Ye_Liu_Han_2022}) often face a trade-off: enhancing privacy typically leads to reduced model utility, and vice versa. Effective defenses against MI attacks must innovate methods that reduce privacy leakage without compromising model utility. Although continued advancements are needed to refine this balance, approaches such as random erasing~\citep{Tran_Nguyen_Mai_Vandierendonck_Cheung_2024}, adversarial noise~\citep{Wen_Yiu_Hui_2021}, and mutual information regularization~\citep{Wang_Zhang_Jia_2021} look promising.

\textbf{Future Research}: To address this challenge, research efforts can be directed at developing adaptive techniques that dynamically adjust privacy-preserving mechanisms based on application-specific requirements. Innovations such as multi-objective optimisation methods would enable simultaneous minimization of privacy leakage and maximization of model utility. Furthermore, collaboration across fields, including deep learning, cryptography, and statistics, will be critical in creating flexible and robust frameworks that ensure privacy without deteriorating model performance.

\subsection{Defining Realistic Threat Models}

Defining realistic threat models for MI attacks is a complex task. As adversarial techniques evolve, traditional threat models often fail to capture the breadth of potential risks, especially in real-world scenarios characterized by noisy data, diverse datasets, and rapidly advancing architectures. One of the primary difficulties in establishing comprehensive threat models is accounting for the adaptability of MI attacks across diverse neural network architectures, such as transformers and multimodal models. Current research shows that MI attacks can exploit architecture-specific vulnerabilities, but the extent of these risks remains underexplored~\citep{Zhang_Liu_Huang_Wang_Lee_Chen_2022, Li_Hao_Wang_Zhu_Wang_Zhang_Feng_2024}. Additionally, most existing threat models are tested under controlled laboratory conditions, which do not always reflect the complexities of real-world deployments. Practical factors, such as heterogeneous datasets, capabilities of the adversary, and model structures, significantly affect the robustness and adaptability of MI attacks~\citep{Ye_Luo_Zhou_Zhu_Shi_Jia_2024, Wang_Zhang_Jia_2021}.

\textbf{Future Research}: It is important for researchers to establish dynamic and flexible frameworks based on realistic threat models. Such frameworks should incorporate: 
(1) diverse attack vectors, including those exploiting new architectures like transformers and GANs~\citep{Chen_Jia_Qi_2020}; (2) robust testing against real-world conditions, such as datasets that are noisier and more heterogeneous than commonly used ones (e.g., CelebA~\citep{Peng_Liu_Zhang_Lan_Ye_Liu_Han_2022}) for MI attack evaluations; and (3) development of defensive strategies that balance model utility and privacy under evolving threat scenarios~\citep{Wen_Yiu_Hui_2021, Tran_Nguyen_Mai_Vandierendonck_Cheung_2024}. 

\subsection{Lack of Scalability and Generalizability of Defenses}
Defensive strategies against MI attacks often struggle with scalability and generalizability, particularly in complex, large-scale systems. Many existing MI defenses, such as differential privacy and homomorphic encryption, are effective in controlled scenarios but face challenges when applied to large-scale datasets or sophisticated architectures. These approaches often require substantial computational resources, which limit their applicability in real-time or resource-constrained environments~\citep{Wang_Si_Wu_2015}.

Another critical issue lies in the specificity of defenses, which are frequently tailored to address vulnerabilities in particular architectures or datasets. For example, defenses such as mutual information regularization are effective for certain neural networks but may fail to generalize to architectures like transformers or graph neural networks. Similarly, the implementation of differential privacy often leads to degraded model utility when extended to large-scale datasets, making it less feasible for practical applications~\citep{Wang_Zhang_Jia_2021, Peng_Liu_Zhang_Lan_Ye_Liu_Han_2022}.

\textbf{Future Research}: To address the challenge of the lack of scalability and generalizability, researchers need to explore more adaptable mechanisms~\citep{Ho_Hao_Chandrasegaran_Nguyen_Cheung_2024}, such as TL-based defenses and adversarial training, which aim to generalize across architectures and domains while retaining performance~\citep{Wen_Yiu_Hui_2021}. However, achieving scalable and generalizable defenses that can be applied to large-scale systems without significant computational trade-offs is a challenging issue, necessitating further investigation into lightweight, architecture-independent approaches.

\subsection{Domain-specific Challenges}
MI attacks pose great risks across many application domains, demanding tailored defensive strategies to mitigate their impact. For example, in healthcare, deep learning models are often trained on sensitive data in diverse formats, such as electronic health records in tabular form and medical X-ray images. While MI attacks need to adapt to these varying data formats, effective defenses in this domain must also accommodate such diversity. At the same time, it is necessary to strike a balance between preserving privacy and ensuring accurate diagnostics and treatment recommendations, thereby safeguarding patient information~\citep{Kaissis_Ziller_Palmbach_Ryffel_Usynin_Trask_LimaJr_Mancuso_Jungmann_Steinborn_2021}.

IoT and smart systems represent another domain where MI attacks can exploit computationally constrained edge devices. These systems, such as smart home devices or industrial IoT platforms, often lack the computational capacity for traditional, resource-intensive defenses. Lightweight encryption or pruning methods can help mitigate these vulnerabilities, but they must be adapted to the unique resource and latency requirements of edge computing~\citep{Alhalabi_2023}.
Given these constraints, conventional encryption techniques may be impractical, increasing the susceptibility of these devices to inference attacks. To address this, lightweight encryption schemes and optimized security protocols are essential, as they can enhance data protection while ensuring efficient transmission and real-time responsiveness~\citep{Al_Hejri_Azzedin_Almuhammadi_Eltoweissy_2024}.

\textbf{Future Research}: To tackle these challenges, domain-specific defense schemes must be developed. These schemes should involve privacy-preserving techniques supported by an understanding of varying needs (e.g., different data types or resource-limited devices) in each field, ensuring both robustness against MI attacks and usability in real-world scenarios. Focusing on tailored approaches will make future studies  better align the goals of privacy, utility, and practicality in diverse application domains.
In the case of IoT and edge computing, further exploration of lightweight cryptographic methods, secure data transmission protocols, and scalable privacy-preserving techniques is necessary to minimize vulnerabilities without overburdening resource-limited devices~\citep{Singh_Sharma_Moon_Park_2024}, especially when attaching AI accelerators to these devices are impractical. Additionally, interdisciplinary efforts that integrate security, privacy, and real-world usability will be critical in ensuring that MI defenses align with the practical needs of different applications.

\subsection{Lack of Standardized Metrics, Diverse Datasets and Open-course Repositories for MI Attacks and Defenses}
The lack of standardized metrics for evaluating MI attacks and defenses continues to impede progress in the field of MI-related research. Establishing clear and consistent metrics for assessing data leakage, attack success rates, and defense robustness is essential for fair comparisons across research works. Existing evaluation methods vary drastically, as shown in Section~\ref{sec:Evaluation Metrics}, making itchallenging to identify universally effective solutions. For instance, the fidelity-related metrics (e.g., MSE and PSNR) of reconstructed data are often employed to assess MI attack success, but these metrics lack consistency in their implementation across studies~\citep{Fredrikson_Jha_Ristenpart_2015}.

The development of representative and diverse benchmark datasets is also important for testing MI defenses under realistic conditions. Publicly available datasets, such as CIFAR-10 or MNIST, are frequently used in MI research but fail to capture the complexities of real-world situations, such as those in healthcare or finance, where representative datasets are lacking, as shown in Section~\ref{sec:Datasets}. Developing domain-specific datasets and evaluation protocols tailored to sensitive applications would significantly enhance the relevance of MI research~\citep{Huang_Gupta_Song_Li_Arora_2021}.

In addition, the adoption of open-source repositories for evaluating MI attacks and defenses can facilitate reproducibility and collaboration within the research community. Such repositories would allow researchers to compare methods directly and transparently, leading to more robust and innovative solutions. Recent efforts, such as those incorporating FL and differential privacy techniques, highlight the importance of comprehensive testing environments for understanding trade-offs between privacy and utility~\citep{Zhou_Zhu_Yu_Li_Peng_Liu_Han_2024}.

\textbf{Future Research}: Researchers should prioritize the design of standardized metrics to quantify data leakage and assess defense robustness consistently across studies~\citep{Qiu_Yu_Fang_Yu_Chen_Wang_Xia_Xu_2024}. Creating representative domain-specific datasets, alongside synthetic datasets for controlled testing, will enhance the realism and reproducibility of evaluations~\citep{Huang_Gupta_Song_Li_Arora_2021}. Although a few open-source repositories exist~\citep{Zhou_Zhu_Yu_Li_Peng_Liu_Han_2024}, each has its own limitation. There is still a need for more open-source repositories and simulation environments to enable transparent and collaborative testing of MI defenses under diverse conditions. One of the main contributions of this work is the development of a comprehensive repository of state-of-the-art research articles, datasets, evaluation metrics, and other essential resources to address this need.

\subsection{Legal and Ethical Considerations}
MI attacks pose profound legal and ethical challenges that demand interdisciplinary solutions. Compliance with global privacy regulations such as GDPR, CCPA, and HIPAA is imperative to address these challenges~\citep{Veale_Binns_Edwards_2018}. These regulations require that defensive strategies not only mitigate risks of data reconstruction but also align with the principles of lawful and ethical data handling, ensuring individuals' privacy rights are preserved~\citep{Nguyen_2024}. Moreover, ethical AI development necessitates balancing robust data protection with transparency, accountability, and fairness. This balance is particularly crucial in sensitive domains like healthcare, where safeguarding patient data must coexist with advancing diagnostic tools~\citep{Thapa_2024}.

Another pressing concern is the potential misuse of MI-related research by malicious actors, emphasizing the need for responsible dissemination of findings. This requires setting clear guidelines on publishing sensitive results to avoid enabling adversarial exploitation while maintaining the spirit of open research~\citep{Hung_2023}. Additionally, collaboration between technical, legal, and ethical experts is crucial to develop frameworks that holistically address these challenges. Initiatives that integrate privacy-preserving AI technologies, compliance checks, and ethical design principles can help produce socially responsible solutions for MI defense~\citep{Makhdoom_Abolhasan_Lipman_Shariati_Franklin_Piccardi_2024}.

\textbf{Future Research}: Research efforts should be devoted to developing frameworks that enable transparency and accountability without compromising data security, particularly in high-stakes fields such as healthcare and finance~\citep{Thapa_2024}. To mitigate the risks of adversarial misuse, researchers must establish guidelines for responsibly disseminating sensitive findings while encouraging open scientific collaboration~\citep{Hung_2023}. Additionally, interdisciplinary collaboration between technologists, ethicists, and legal experts will be essential to advance socially responsible AI development. Such partnerships can support the creation of adaptive and globally relevant frameworks that balance ethical considerations with technical innovation~\citep{Makhdoom_Abolhasan_Lipman_Shariati_Franklin_Piccardi_2024}.

\section{Conclusion}
\label{sec:Conclusion}

This survey has provided a comprehensive overview of MI attacks and corresponding defense mechanisms, offering a structured taxonomy of MI attacks based on diverse techniques and an in-depth review of their applications across key domains such as biometric recognition, healthcare, and finance. Through detailed analysis and discussion, we emphasize the increasing importance of developing robust defenses to protect sensitive data in deep learning systems. While summarizing existing defense strategies, we recognize that many current approaches face limitations in terms of scalability, generalizability, and maintaining model utility. Furthermore, most defenses are evaluated under simplified threat models that may not fully capture real-world adversarial capabilities. These limitations highlight the need for continued research in designing more practical and adaptable solutions.

We also identify several promising directions for future work. These include defining realistic threat models, enhancing the explainability of defense techniques, and addressing domain-specific challenges. By tackling these issues, we can pave the way toward developing more secure deep learning systems that are resilient to MI attacks.

\backmatter



\bmhead{Acknowledgments}
This research was supported in part by an Australia Research Council (ARC) Discovery Project Grant: DP230102828.

\section*{Declarations}

\bmhead{Conflict of interest}
The authors declare that they have no conflict of interest.

\bibliography{sn-bibliography}

\end{document}